\documentclass[12pt,showpacs,pra,superscriptaddress]{revtex4-1}

\usepackage{amssymb}
\usepackage{amsfonts}
\usepackage{amsmath}

\usepackage{dcolumn}
\usepackage{bm}
\usepackage{epsfig}
\usepackage{epsf}
\usepackage[]{graphicx}
\usepackage[sort&compress]{natbib}
\usepackage{color}
\usepackage{xspace}

\usepackage{longtable}

\renewcommand{\i}{{\rm i}}
\renewcommand{\d}{{\rm d}}
\newcommand {\ee}{{\rm e}}

\newcommand {\bfn} {{\bf n}}
\newcommand {\bfp} {{\bf p}}
\newcommand {\bfr} {{\bf r}}
\newcommand {\bfv} {{\bf v}}

\newcommand {\bfR} {{\bf R}}

\newcommand {\calA} {{\cal A}}
\newcommand {\calN} {{\cal N}}
\newcommand {\calU} {{\cal U}}

\newcommand {\E}  {{\varepsilon}}
\newcommand {\om}  {{\omega}}
\newcommand {\Om}  {{\Omega}}

\newcommand {\aTF}  {a_{\rm TF}}

\newcommand {\Ld}  {L_{\rm d}}

\newcommand{\Lch}{{L_{\rm ch}}}

\newcommand{\lamu}{{\lambda_{\rm u}}}
\newcommand{\lamch}{{\lambda_{\rm ch}}}

\begin{document}

\title{Simulation of Channeling and Radiation of 855 MeV Electrons and Positrons 
in a Small-Amplitude Short-Period Bent Crystal}

\author{Andrei V. Korol \footnote{E-mail: korol@mbnexplorer.com}}
\affiliation{MBN Research Center, Altenh\"{o}ferallee 3, 
60438 Frankfurt am Main, Germany}
%
\author{Victor G. Bezchastnov}
\affiliation{A.F. Ioffe Physical-Technical Institute, Politechnicheskaya Str. 26, 
194021 St. Petersburg, Russia}
\affiliation{Peter the Great St. Petersburg Polytechnic 
University, Politechnicheskaya 29, 195251 St. Petersburg, Russia.
}
\author{Gennady B. Sushko} 
\affiliation{MBN Research Center, Altenh\"{o}ferallee 3, 
60438 Frankfurt am Main, Germany}
\author{Andrey V. Solov'yov}
\affiliation{MBN Research Center, Altenh\"{o}ferallee 3, 
60438 Frankfurt am Main, Germany}

%


\begin{abstract}
Channeling and radiation are studied for the relativistic electrons and 
positrons passing through a Si crystal periodically bent with a small 
amplitude and a short period. 
Comprehensive analysis of the channeling 
process for various bending amplitudes is presented on the grounds 
of numerical simulations. 
The features of the channeling are highlighted 
and elucidated within an analytically developed continuous potential 
approximation. 
The radiation spectra are computed and discussed.
\end{abstract}

\maketitle
\section{Introduction \label{Introduction}}

Channeling of the charged projectiles in crystals establishes a field of 
research important both with respect to the fundamental theoretical studies as well as to 
the ongoing experiments (see, e.g., the monograph~\cite{ChannelingBook2014} and the 
references therein). Starting from the first theoretical predictions by 
Lindhard~\cite{Lindhard1965}, the channeling is known to occur when the projectiles move 
in the crystals preferably along the crystalline planes or axes. This phenomenon opens a 
possibility to manipulate the beams of projectiles, in particular by 
deflecting them in bent crystals. 

Of particular interest for the theory and applications is the radiation produced by the 
channeling projectiles. The radiation spectra display distinct lines related to 
the inter-planar oscillations or axial rotations of the particles involved into 
the planar or axial channeling, respectively, in contrast to the bremsstrahlung-type 
spectra produced by the non-channeling projectiles. In periodically bent crystals, 
the channeling radiation exhibit additional {\em undulator lines} as a result of modulation 
of the transverse velocity of the moving particles by the bent crystalline structure. 
The positions of the undulator lines depend on the bending period and the projectile energies 
and thereby can be tuned in an experiment. Theoretically, a crystalline undulator has been 
suggested~\cite{KSG1998,KSG_review_1999} as a source of the monochromatic radiation of sub-MeV 
to MeV energies, and different experiments have been being performed to produce and detect 
the undulator radiation. 

The original concept of the crystalline undulator assumes the projectiles to move through 
the crystal following the periodically bent crystalline planes or axes. For such motions, 
the undulator modulation frequencies are smaller than the frequencies of the 
channeling oscillations or rotations providing the undulator spectral lines to arise 
at the energies below the energies of the channeling lines. Recently, channeling 
has been studied for periodically bent crystals with the bending period shorter 
than the period of channeling oscillations in the straight 
crystal~\cite{Kostyuk_PRL2013,Multi_GeV_2014}. The emergent radiation was shown to display
spectral lines at the energies exceeding the energies of the channeling peaks. 
The corresponding theoretical simulations have been performed for the electron and positron 
channeling in the silicon crystals with the (110)-planes bent according to the shapes 
\begin{equation}
\delta(z) = a \cos (2\pi z / \lamu), 
\label{PB_profile:eq}
\end{equation}
where $z$ is a coordinate along the planes in the straight 
crystal, whereas $a$ and $\lamu$ are the bending amplitude and period, respectively. 
The simulations have also assumed the amplitude $a$ to be substantially smaller than the 
inter-planner distance $d$ in the straight crystal. In contrast to the channeling in 
conventional crystalline undulator, for the newly proposed crystalline structures 
the channeling projectiles do not follow the short-period bent planes. The positrons 
move bouncing between the bent planes, whereas the electrons move preferable through 
the small-amplitude ``waves'' of the bent structures. Yet the motion of the projectiles 
acquires a short-period modulation resulting from the bending and explicitly seen in the 
theoretically simulated  trajectories~\cite{Multi_GeV_2014}. These modulations are of a 
regular jitter-type and responsible for producing the high-energy monochromatic 
radiation. Interestingly, a similar radiative mechanism has recently been studied with 
respect to the radiation produced by relativistic particles in interstellar environments 
with turbulent small-scale fluctuations of the magnetic 
field~\cite{Medvedev_2000,Kellner_2013}.

It requires, in our view, some future clarifications on whether the jitter-type 
channeling motions in crystals provide observable properties (in particular for the 
radiation detected) inherent to the conventional undulator. Within the current studies, 
we adopt the term {\em small-amplitude short-period} (SASP) to designate the bent 
crystalline structures suggested in Ref.~\cite{Kostyuk_PRL2013}. Important is that, the 
SASP bent crystals can be regarded as possible efficient sources of tunable monochromatic 
hard-energy radiation and therefore are of immediate interest for applications and 
further theoretical studies. 

Since introducing the SASP bent crystals~\cite{Kostyuk_PRL2013}, several 
theoretical and experimental studies have already been performed on the new regime of 
channeling and the radiation produced~\cite{Backe_EtAl_PRL2014,Multi_GeV_2014}. However, 
further efforts are certainly required for more detailed investigations on statistical 
properties of channeling as well as on computing the radiation spectra for a variety of 
conditions including different bending parameters, lengths of the crystalline samples, 
beam energies and angular apertures for detecting the radiation. The focus of the present 
paper is on the theoretical simulations for the electrons and positrons with the 
energies of $855$~MeV that are the beam energies achievable in experiments at Mainz 
Microtron Facility~\cite{Backe_EtAl_PRL2014,Backe_EtAl_PRL_115_025504_2015}. 

As in a number of our recent studies, three-dimensional simulations of the 
propagation of ultra-relativistic projectiles through the crystal are performed 
by using the \textsc{MBN Explorer} package~\cite{MBN_ExplorerPaper,MBN_ExplorerSite}.  
This package was originally developed as a universal numerical tool to study structure and 
dynamics on the spatial scales from nanometers and above for a wide range of complex atomic 
and molecular systems. 
In order to address the channeling phenomena, an additional module has been incorporated 
into the \textsc{MBN Explorer} to compute the motion for relativistic projectiles along with 
dynamical simulations of the propagation environments, including the crystalline structures, 
in the course of projectile's motion~\cite{ChanModuleMBN_2013}. These computations advance to 
account for the interaction of the projectiles with the separate atoms of the environments, 
whereas a variety of interatomic potentials implemented in~\textsc{MBN Explorer} support 
rigorous simulations of various media. The developed software package can be regarded as a 
powerful numerical tool to uncover the dynamics of relativistic projectiles in crystals, 
amorphous bodies, as well as in biological environments. Its efficiency and reliability has 
already been benchmarked for the channeling in 
crystals~\cite{ChanModuleMBN_2013,BentSilicon_2013,Sub_GeV_2013,BentSilicon_2014,%
Multi_GeV_2014,Sushko_EtAl_NIMB_v355_p39_2015}. 
The calculated with the \textsc{MBN Explorer} relativistic motion, represented by the 
projectile's coordinates and velocities at the instances of the propagation time, is used 
as the input data to compute the spectral and/or the spectral-angular distributions of the 
emitted radiation. A module for calculating the radiation emergent from the channeling 
is included into the latest version of \textsc{MBN Explorer}~\cite{MBN_ExplorerSite}.

\section{Theoretical Framework \label{Approach}}

We consider the motion of the projectiles through the crystal 
as being governed by the laws of classical relativistic dynamics: 
\begin{eqnarray}
\partial \bfr / \partial t = \bfv,
\qquad
\partial \bfp / \partial t = - q \, \partial U/\partial \bfr,
\label{MC_Simulations.01}
\end{eqnarray}
where $\bfr(t)$ is the coordinate, $\bfv(t)$ is the velocity and $\bfp(t) = m\gamma\bfv(t)$ 
is the momentum of the particle at the propagation time $t$,  
$\gamma = \left(1-v^2/c^2\right)^{-1/2} = \E/mc^2$ is the relativistic Lorentz-factor, $\E$ 
and $m$ are the energy and the rest mass of the particle, respectively, and $c$ is the speed 
of light. The driving force, sensitive to the charge $q$ of the projectile, stems from an 
electrostatic potential $U=U({\bf r})$ for the particle-crystal interaction. 
\textsc{MBN Explorer} allows for computing the latter potential from the 
interactions of the projectile with the individual atoms,  
\begin{eqnarray} 
U(\bfr) = \sum_{j} U_{\mathrm{at}}\left(\left|\bfr - \bfR_j\right|\right). 
\label{MC_Simulations.02}
\end{eqnarray} 
The spacial locations $\bfR_j$ of the atoms are selected according to the crystalline 
structures of the interest (including the effect of thermal fluctuations). 
The atomic potentials $U_{\mathrm{at}}$ in our simulations are evaluated within 
the Moliere approximation \cite{Moliere}. A rapid decrease of these potentials with 
increasing the distances from the atoms allows to truncate the 
sum (\ref{MC_Simulations.02}) in practical calculations. \textsc{MBN Explorer} provides 
an option to restrict the atoms contributing to the interaction $U({\bf r})$ to these 
located inside a cut-off sphere around the coordinate $\bfr$, as well as invokes an 
efficient linked-cell algorithm to search for the atoms inside this sphere.   

A particular feature of \textsc{MBN Explorer} is simulating the environment ``on the fly'' 
i.e. in the course of computing the motion of the projectiles. For channeling, the crystalline 
lattice is simulated inside a box surrounding the position of the projectile and the simulation 
box shifts accordingly to the motion of the projectile. The coordinate frame for the 
simulations has the $z$-axis along the incoming beam and parallel to
the crystalline planes responsible for the channeling, whereas the $y$ axis is set 
perpendicular to these planes. In order to exclude an accidental axial channeling, 
the $z$-axis should avoid major crystallographic directions. For channeling along 
the (110) planes, we opted to define the $z$-axis by a direction $\langle -n n m \rangle$ with 
$n\gg m \sim 1$. 

The simulation box is filled by the crystalline lattice with the nodes 
$\bfR_j = \bfR_j^{(0)} + \boldsymbol{\Delta}_j$, 
selected with account for the thermal displacements $\boldsymbol{\Delta}_j$ of the 
atomic nuclei with respect to the equilibrium positions $\bfR_j^{(0)}$. 
The equilibrium nodes correspond to the Bravais cells for the crystal, whereas 
the Cartesian components $\Delta_{jk}$ ($k=x,y,z$) of the thermal displacements are 
selected randomly according to the normal distribution 
\begin{eqnarray} 
w(\Delta_{jk})=
\frac{1}{\sqrt{2\pi u_T^2}} \exp\left(- {\Delta_{jk}^2 \over 2u_T^2}\right)\,.
\label{MC_Simulations.05}
\end{eqnarray}
The values of the amplitude $u_T$ of the thermal vibrations are well-known for 
various crystals and can be found in Ref.~\cite{Gemmel}. For the silicon crystals 
at room temperature we use the value $u_T = 0.075$ \AA.

The equations (\ref{MC_Simulations.01}) are numerically integrated forth from $t=0$ 
when the particle enters the crystal at $z=0$. The initial values $x_0$ and $y_0$ for the 
transverse coordinates are selected randomly from an entrance domain where the beam can be 
guided by the crystalline planes to get into the inside of the crystal. The size of this 
domain is taken between $2d$ and $5d$ in the $x$-direction and between $d$ and 
$3d$ in the $y$-direction, where $d$ is the inter-planner separation for the crystal.
The initial velocity $\bfv_0=(v_{0x},v_{0y},v_{0z})$ has the value determined by the 
beam energy and is predominantly oriented in the $z$-direction, i.e. the values $v_{0x}$ and 
$v_{0y}$ are small compared to the value $v_{0z} \approx c$. The non-vanishing $v_{0x}$ and 
$v_{0y}$ can be adjusted to account for the beam emittance. The data presented below in 
Sects. \ref{Lengths} and \ref{Spectra} are obtained for the zero emittance, $v_{0x}=v_{0y}=0$.

When computing the motion of the projectile through the crystal with \textsc{MBN Explorer}, 
an efficient algorithm of ``dynamic simulation box'' \cite{ChanModuleMBN_2013} is used as 
follows. Inside a simulation box, the particle interacts with the atoms of the cutoff sphere. 
As the particle moves, the sphere shifts and at some point meets an edge of the box. 
Once this happens, a new simulation box is introduced being centered at the current position 
of the projectile. The new box is then filled with the crystalline lattice such that the nodes 
inside the intersection of the old and the new simulation boxes remain unchanged 
i.e. not being simulated anew with including the effect of the thermal vibrations. 
This allows to avoid a spurious change in the driving force as well as reduces 
numerical efforts in simulating the crystalline environment in the course of the 
projectile's motion. 

The above described numerical propagation procedure terminates when the $z$-coordinate of the 
projectile approaches the thickness $L$ of the crystal under study. To simulate a periodically 
bent crystal, the $y$-coordinates for each lattice node $\bfR_j=(X_j,Y_j,Z_j)$ are obtained 
from these for the straight crystal by the transformation $Y_j \to Y_j + \delta(Z_j)$ 
with $\delta(Z_j)$ being determined by the bending profile~(\ref{PB_profile:eq}).

Employing the Monte-Carlo technique for sampling the incoming projectiles as 
well as for accounting for the thermal fluctuations of the crystalline lattice 
yields a statistical ensemble of trajectories simulated with \textsc{MBN Explorer}. 
The trajectories then can be used for computing the radiation from the projectiles 
passing through the crystal. The energy emitted per unit frequency $\omega$ 
within the cone $\theta\leq \theta_{\max}$ along the $z$ axis is computed as 
follows 
\begin{eqnarray}
{\d E(\theta\leq\theta_{\max}) \over \d \om}
=
{1 \over N_0}
\sum_{n=1}^{N_0} 
\int\limits_{0}^{2\pi}
\d \phi
\int\limits_{0}^{\theta_{\max}}
\theta \d\theta\,
{\d^2 E_n \over \d \om\, \d\Om}, 
\label{MC_Simulations.06}
\end{eqnarray}
where the sum is carried over the simulated trajectories of the total number $N_0$, 
$\Omega$ is the solid angle corresponding to the emission angles $\theta$ and $\phi$, 
and $\d^2 E_n/\d \om\, \d\Om$ is the energy per unit frequency and unit solid angle 
emitted by the projectile moving along the $n$th trajectory. The general equation 
(\ref{MC_Simulations.06}) accounts for contributions of {\em all the segments} of 
simulated trajectories i.e. the segments of the channeling motion as well as the 
segments of motion out of the channeling regime.

The radiation emitted by the individual projectiles is computed within the 
quasi-classical approach developed by Baier and Katkov. 
For the details of the formalism as well as various applications to radiative 
processes we refer to the monograph~\cite{Baier} (we also mention Appendix A in 
Ref. \cite{Uggerhoj2011} where a number of intermediate steps of the formalism 
are evaluated explicitly). 
It is to be pointed out that, along with the classical description of the motion of 
projectiles, the formalism includes the effect of quantum radiative recoil i.e. it accounts for 
the change of the projectile energy due to the photon emission. The impact of 
the recoil on the radiation is governed by the ratio $\hbar \om /\E$, and 
the limit $\hbar \om /\E \to 0$ corresponds to purely classical description 
of the radiative process. For channeling, the classical framework is found to 
be adequate for calculations of the radiation produced in straight, bent and 
periodically bent crystals by the electrons and positrons with the sub-GeV 
beam energies (see, e.g, the monograph~\cite{ChannelingBook2014} and the 
references therein). For higher beam energies, the quantum recoil can become 
important, and has recently be shown~\cite{Multi_GeV_2014} to significantly 
influence the radiation produced in SASP bent crystals by the multi-GeV 
electrons and positrons.

To compute the spectral intensity of the radiation produced by the particle 
moving along the trajectory $\bfr=\bfr(t)$ in the crystal of the thickness $L$, 
we use the equation~\cite{ChanModuleMBN_2013,ChannelingBook2014}:
\begin{eqnarray}
{ \d^2 E \over \hbar\d\om\, \d \Om} =
\alpha q^2\omega^2\,{ (1+u) (1 + w)\over 4 \pi^2 }
\Bigg[
&\,&
{w \left|I_z\right|^2\over \gamma^2 (1 + w)}
+
\left| \sin\phi  I_{x} - \cos\phi I_{y}\right|^2
\nonumber \\
&\,&
+
\left|
\theta I_z
-
 \cos\phi  I_{x}
- \sin\phi  I_{y}
\right|^2
\Bigg],
\label{wkb:eq.04}
\end{eqnarray}
where $\alpha$ is the fine structure constant, $q$ is the charge of the projectile in 
units of the elementary charge, $u=\hbar\om/(\E-\hbar\om)$ and $w=u^2/(1+u)$. 
The quantities $I_{x,y,z}$ involve the integrals with the phase functions, 
\begin{eqnarray}
\begin{array}{l}
\displaystyle
I_{z}
=
\int\limits_0^{\tau} \d  t\,  \ee^{\i \om^{\prime} \Phi(t)}
-
{\i \over \om^{\prime}}
\left(
{\ee^{\i\, \om^{\prime}  \Phi(0) } \over D(0)}
-
{\ee^{\i\,\om^{\prime} \Phi(\tau)} \over D(\tau)}
\right)\,,
\\
\displaystyle
I_{x,y}
=
\int\limits_0^{\tau} \d  t\,
{ v_{x,y} (t) \over c}\, \ee^{\i \om^{\prime} \Phi(t)}
-
{\i \over \om^{\prime}}
\left(
{ v_{x,y} (0) \over c}
{\ee^{\i\, \om^{\prime}  \Phi(0) } \over D(0)}
-
{ v_{x,y} (\tau) \over c}
{\ee^{\i\,\om^{\prime} \Phi(\tau)} \over D(\tau)}
\right)\,,
\end{array}
\label{Section3.5_2:eq.02}
\end{eqnarray}
where $\tau=L/c$ is the time of flight through the crystal, 
$D(t) = 1 - \bfn\cdot\bfv(t)/c$, $\Phi(t) = t - \bfn\cdot\bfr(t)/c$, 
$\bfn$ is the unit vector in the emission direction, 
and $\om^{\prime}=(1+u)\om$. The quantum radiative recoil is accounted for by 
the parameter $u$ which vanishes in the classical limit. 

\section{Channeling and Radiation for $855$~MeV Electrons and Positrons}

Within the above described theoretical framework and by exploiting the 
\textsc{MBN Explorer} package, the trajectories and radiation spectra 
have been simulated for the $\E=855$ MeV electrons and positrons. The 
projectiles were selected as incoming along the (110) crystallographic 
planes in a straight silicon crystal, and the SASP bent crystalline 
profiles were introduced according to Eq.~(\ref{PB_profile:eq}). 

The amplitude and period of bending were varied within the intervals 
$a=0.05 \dots 0.95$ \AA{} (providing $a$ to be smaller than the distance 
$d=1.92$ \AA{} between the (110) planes in the straight Si crystal) and 
$\lamu = 100 \dots 2500$ nm. 
These ranges include the values used in the recent 
theoretical \cite{Kostyuk_PRL2013,Multi_GeV_2014} and experimental 
\cite{Backe_EtAl_PRL2014} studies 
on the channeling and radiation for SASP bending.  
The results we have obtained reveal a variety of peculiar features of the channeling 
in SASP bent Si(110). A qualitative analysis of these results is presented in 
Section \ref{Lengths}.

The calculated emission spectra cover a wide range of the photon energies, 
from $\lesssim 1$~MeV up to $40$~MeV. The calculations were performed for the crystal 
thicknesses $12$, $25$, $75$ and $150$ micron measured along the beam direction. 
The integration over the emission angle $\theta$ was carried out for two particular 
detector apertures determined by the values $\theta_{\max}=0.21$ and $4$~mrad. 
The first value is close to one used in the experiments with the $855$ MeV electron 
beam at Mainz Microtron facility 
\cite{Backe_EtAl_2008,Backe_EtAl_2011,BackeLauth_2013,Backe_EtAl_PRL2014,%
Backe_EtAl_PRL_115_025504_2015}, 
and is much smaller than the natural emission angle for the beam energy, 
$\gamma^{-1} \approx 0.6$ mrad. 
Therefore, the corresponding spectra refer to a nearly forward emission. 
The second angle value, in contrast, significantly exceeds 
the value $\gamma^{-1}$ providing the emission cone with $\theta \leq \theta_{\max}$ 
to collect almost all the radiation from the relativistic projectiles. The latter 
situation corresponds to the conditions at the experimental setup at 
SLAC~\cite{Wienands_EtAl_PRL_v114_074801_2015}. 
The discussion of the calculated emission spectra is given in 
Section \ref{Spectra}.  

\subsection{Statistical Properties of Channeling \label{Lengths}}

For each type of the projectiles and different sets of the bending amplitudes and periods 
(including the case of the straight crystal with $a=0$), the numbers $N_0$ of the 
simulated trajectories were sufficiently large (between $4000$ and $7000$) thus enabling 
a reliable statistical quantification of the channeling process. Below we define and 
describe the quantities obtained. 

A randomization of the ``entrance conditions'' for the projectiles (in particular sampling 
the entrance locations as explained in Sect. \ref{Approach}) makes the different projectiles 
to encounter differently scattering with the crystalline atoms upon entering the crystal. 
As a result, not all the simulated projectiles start moving through the crystal in a 
channeling mode. 
A commonly used parameter to quantify the latter property is 
{\em acceptance} defined as the ratio $\calA = N_{\mathrm{acc}}/N_0$ of the number 
$N_{\mathrm{acc}}$ of particles captured into the channeling mode once entering the 
crystal (the accepted particles) to the number $N_0$ of the incident particles. 
The non-accepted particles experience over-barrier motion unrestricted in the inter-planar 
directions.

It is order to notice that different theoretical approaches to the interaction of the 
projectiles with the crystalline environments are also different in criteria 
distinguishing between the channeling and the over-barrier motions. For example, the 
continuous potential approximation \cite{Lindhard1965} decouples the transverse 
(inter-planar) and longitudinal motions of the projectiles. 
As a result, it is straightforward to distinguish the channeling projectiles as those 
with transverse energies not exceeding the height of the inter-planar potential barrier. 
In our simulations, the projectiles interact, as in reality, with the individual atoms of the 
crystal. 
The inter-planar potential experienced by the projectiles vary rapidly in the 
course of their motion and couples the transverse and longitudinal degrees of freedom. 
Therefore, another criteria are required to select the channeling episodes in the 
projectile's motion. 
We assume the channeling to occur when a projectile, while moving 
in the same channel, changes the sign of the transverse velocity $v_y$ at least two 
times \cite{KKSG_simulation_straight}. 
The latter criteria have also to be supplemented by geometrical definitions of the 
crystalline channels. 
Here, we refer to the straight crystals where the positron channels are the volume areas 
restricted by the neighboring (110) planes, whereas the electron channels are the areas 
between the corresponding neighboring mid-planes. 
For SASP bent crystals, the simulations show that, at small values of the bending amplitude, 
the channeling occur in the above defined channels for the straight crystal. 
We will refer to such situation as to the ``conventional'' channeling.
 With the amplitudes increasing above some values, the positrons 
tend to channel in the electron channels for the straight crystal, and vice versa, the 
electrons become channeling in the positron channels. 
The latter situation will be regarded as the ``complementary'' channeling. 
We will provide below a quantification for both channeling regimes. 

For an accepted projectile, the channeling episode lasts until an event of 
de-channeling when the projectile leaves the channel. The de-channeling events 
occur mostly as cumulative outcomes of multiple collisions of the projectiles 
with the crystalline constituents. Also, a rare large-angle scattering 
collision can lead to de-channeling. De-channeling of the accepted particles 
is conveniently quantified by the {\em penetration depth} $L_{\mathrm{p1}}$ 
evaluated as the mean longitudinal extension of the primary channeling segments 
of the trajectories. The latter segments are those that start at the crystalline 
entrance and end at the first de-channeling event experienced by the particle 
inside the crystal \cite{ChanModuleMBN_2013}. The penetration depth can be 
related to a commonly used  de-channeling length $\Ld$. Within the 
framework of the diffusion theory of de-channeling 
(see, e.g., Ref.~\cite{BiryukovChesnokovKotovBook}), the fraction of the 
channeling particles at large distances $z$ from the entrance decays 
exponentially, $\propto \exp(-z/\Ld)$, 
with increasing $z$ \cite{footnote}. 
Assuming the decay law to be applicable 
for all $z$, the penetration depth can be evaluated as the integral 
$L_{\mathrm{p1}} = \int_0^L (z/\Ld) \exp(-z/\Ld) \d z$ and appears to be smaller 
than $\Ld$ for the finite lengths $L$ of the crystalline samples. For sufficiently 
long crystals, $L \gg \Ld$, the penetration depth approaches the de-channeling length, 
$L_{\mathrm{p1}} \to \Ld$.

Random scattering of the projectiles on the crystalline constituents can also result 
in a {\em re-channeling} process of capturing the over-barrier particles into 
the channeling mode of motion. 
This process is quantified by the {\em re-channeling length}, 
$L_{\rm rech}$, defined as an average distance along the crystal between the trajectory 
points corresponding to the successive events of de-channeling and re-channeling. 
We notice that, the re-channeling events are more frequent (and, correspondingly, the 
re-channeling lengths are shorter) for the negatively charged projectiles than these 
for the positively charged ones. 
For a qualitative explanation of this feature we refer to 
Ref.~\cite{KKSG_simulation_straight}.

In a sufficiently long crystal, the projectiles can experience de-channeling and 
re-channeling several times in the course of propagation. 
These multiple events are 
accounted for by an additional pair of lengths, the penetration length 
$L_{\mathrm p2}$ and the total channeling length $\Lch$ that characterize the 
channeling process in the whole crystal. 
The depth $L_{\mathrm p2}$ is the average 
extension calculated with respect to {\em all} channeling segments in the trajectories, 
i.e. not only with respect to the segments that start from the crystalline entrance 
and are used to evaluate $L_{\mathrm p1}$ but also including the segments that appear 
inside the crystal due to the re-channeling events followed by de-channeling. 
The total channeling length $\Lch$ is computed as the average length with respect to 
all channeling segments per trajectory. 

As a result of re-channeling, the projectiles become captured into the channels when 
possess, statistically, the incident angles of the values smaller than the Lindhard's 
critical angle values $\Theta_{\mathrm L}$ \cite{Lindhard1965}. 
Therefore, for sufficiently thick crystals, the lengths $L_{\mathrm p2}$ provide the estimates 
of the de-channeling lengths $\Ld$ for the beams with the emittance values 
$\approx \Theta_{\mathrm L}$.

The above described statistical quantities have been calculated from the 
simulations on the channeling for the 855 MeV electrons and positrons in the straight 
and SASP bent Si(110) crystals of the thickness $L=150$~micron. The value of the 
bending period has been fixed at $\lamu=400$~nm, and different values of the 
bending amplitude $a$ have been selected. The results on the acceptance and 
characteristic lengths are presented in Table~\ref{Table_ep-lamu-fixed}. All 
the data refer to the zero emittance beams. Statistical uncertainties due to the 
finite (but sufficiently big) numbers of the simulated trajectories correspond 
to the confidence probability value of $0.999$.

\begingroup
\squeezetable
\begin{table}
\caption{
Acceptance ${\cal A}$, penetration depths $L_{\mathrm p1,2}$, total channeling length 
$L_{\rm ch}$ and re-channeling length $L_{\rm rech}$ for $855$ MeV electrons and 
positrons channeling in $L=150~\mu$m thick straight and SASP bent Si(110). 
The bending period is $\lamu=400$~nm, 
and the different values of the bending amplitude $a$ are examined. 
Two sets of statistical quantities given in the upper and bottom lines for $a=0.7, 0.8$ and 
0.9~\AA{} are deduced from the ``conventional'' and ``complementary'' trajectories, 
respectively (for the detailed explanations see the text). 
\label{Table_ep-lamu-fixed}}
\begin{ruledtabular}
\begin{tabular}{cccccc}
 & & 
\multicolumn{3}{c}{electron channeling}
&  \\
$a$ (\AA)&$\cal{A}$ & $L_{\rm p1}$ ($\mu$m) & $L_{\rm p2}$ ($\mu$m) 
& $L_{\rm ch}$ ($\mu$m)
& $L_{\rm rech}$ ($\mu$m) \\
\hline
$0.00$ & $0.64$& $ 11.89\pm0.49$& $11.19\pm 0.19$ & $42.10\pm 1.21$& $24.12\pm  0.76$ \\
$0.05$ & $0.61$& $ 9.05\pm0.43$ & $ 8.98\pm 0.17$ & $39.35\pm 1.49$& $21.44\pm  0.87$ \\
$0.10$ & $0.54$& $6.37\pm0.32$  & $6.60\pm 0.10$  & $34.46\pm 1.40$& $19.20\pm  0.83$ \\
$0.15$ & $0.48$& $5.45\pm0.30$  & $5.69\pm 0.09$  & $30.78\pm 1.39$& $19.39\pm  0.89$ \\
$0.20$ & $0.48$& $5.77\pm0.33$  & $5.71\pm 0.10$  & $26.38\pm 1.16$& $23.43\pm  1.01$ \\
$0.25$ & $0.50$& $6.79\pm0.43$  & $6.40\pm 0.14$  & $25.87\pm 1.17$& $26.38\pm  1.15$ \\
$0.30$ & $0.56$& $8.20\pm0.43$  & $7.59\pm 0.17$  & $26.07\pm 1.02$& $30.31\pm  1.12$ \\
$0.40$ & $0.70$& $10.74\pm0.57$ & $10.04\pm 0.26$ & $30.20\pm 1.28$& $32.89\pm  1.40$ \\
$0.50$ & $0.73$& $11.00\pm0.57$ & $10.01\pm 0.26$ & $31.37\pm 1.43$& $30.22\pm  1.40$ \\
$0.60$ & $0.78$& $10.63\pm0.55$ & $10.23\pm 0.27$ & $32.89\pm 1.61$& $29.22\pm  1.49$ \\
$0.70$ & $0.81$& $  8.83\pm0.47$& $  9.09\pm 0.23$& $27.70\pm 1.43$& $31.30\pm  1.62$ \\
       & $0.64$& $  5.25\pm0.25$& $  5.70\pm 0.11$& $20.12\pm 1.10$& $29.51\pm  1.53$ \\
$0.80$ & $0.71$& $  4.71\pm0.23$& $  5.44\pm 0.10$& $21.78\pm 1.08$& $26.33\pm  1.26$ \\
       & $0.79$& $  7.06\pm0.33$& $  7.70\pm 0.16$& $26.03\pm 1.05$& $29.21\pm  1.40$ \\
$0.90$ & $0.36$& $  2.16\pm0.10$& $  2.63\pm 0.03$& $13.55\pm 0.67$& $22.77\pm  1.04$ \\
       & $0.72$& $  8.10\pm0.17$& $  7.86\pm 0.17$& $27.18\pm 1.34$& $28.84\pm  1.30$ \\
\hline
 & & 
\multicolumn{3}{c}{positron channeling}
&  \\
$a$ (\AA)&$\cal{A}$ & $L_{\rm p1}$ ($\mu$m) & $L_{\rm p2}$ ($\mu$m) 
& $L_{\rm ch}$ ($\mu$m)
& $L_{\rm rech}$ ($\mu$m) \\
\hline
$0.00$& $0.95$ & $131.45\pm 3.11$ &$100.03 \pm 4.11$ & $131.19 \pm 3.02$ & $29.75 \pm 4.20$  \\
$0.10$& $0.92$ & $124.39\pm 2.78$ &$  84.45\pm 3.09$ & $122.59 \pm 2.66$ & $29.38 \pm 2.49$  \\
$0.20$& $0.89$ & $115.38\pm 2.99$ &$  65.04\pm 2.62$ & $111.97 \pm 2.81$ & $27.48 \pm 1.87$  \\
$0.30$& $0.88$ & $  99.46\pm 3.44$ &$  48.34\pm 2.25$ & $100.63 \pm 3.04$ & $25.82 \pm 1.59$  \\
$0.40$& $0.86$ & $  84.68\pm 4.05$ &$  36.65\pm 2.12$ &$  87.50 \pm 3.53$ & $26.23 \pm 1.64$  \\
$0.50$& $0.84$ & $  63.85\pm 3.94$ &$  23.44\pm 1.47$ &$  70.68 \pm 3.33$ & $24.21 \pm 1.30$ \\
$0.60$& $0.80$ & $  41.70\pm 3.23$ &$  13.76\pm 0.84$ &$  52.96 \pm 2.64$ & $22.26 \pm 1.00$ \\
$0.70$& $0.74$ & $  13.41\pm 1.56$ &$    6.23\pm 0.29$ &$  31.53 \pm 1.58$ & $20.22 \pm 0.94$ \\
      & $0.50$ & $    6.12\pm 0.34$&$    3.44\pm 0.09$ &$  22.17 \pm 1.03$ & $16.32 \pm 0.75$ \\
$0.80$& $0.58$ & $    3.60\pm 0.19$&$    3.60\pm 0.04$ &$  20.77 \pm 0.77$ & $19.17 \pm 0.72$ \\
      & $0.66$ & $    9.20\pm 0.43$&$    6.10\pm 0.15$ &$  27.75 \pm 1.01$ & $21.66 \pm 0.84$ \\
$0.90$& $0.29$ & $    2.01\pm 0.09$&$    2.78\pm 0.02$ &$  17.74 \pm 0.77$ & $18.40 \pm 0.73$ \\
      & $0.80$ & $  15.94\pm 0.76$ &$  12.50 \pm 0.34$ &$  40.68 \pm 1.56$ & $27.26 \pm 1.14$ \\
\end{tabular}
\end{ruledtabular}
\end{table}
\endgroup


Let us first discuss the channeling properties for small to moderate values of the 
bending amplitude, $a\leq 0.6$ \AA{}. 
Comparing the values of $L_{\mathrm p1}$, 
$L_{\mathrm p2}$ and $\Lch$ for the electrons with these for the positrons reveals 
that, all three lengths for the positrons noticeably exceed, 
up to the order of magnitude, the corresponding lengths for the electrons. 
The latter is not surprising given that fact that the channels 
optimally guiding the particles through the crystal are 
different for the positrons and electrons due to the different character of the 
the particle-crystal interactions. 
As the interactions with the crystalline atoms are repulsive for the positrons and attractive 
for the electrons, the channels guide these 
particles to move in the domains with low and high content of the crystalline atoms, 
respectively. 
In course of the projectile's motion, the electrons experience scattering on 
the atoms more frequent as the positrons. 
As a result, for the electrons the 
de-channeling process develops faster and results in shorter penetration and channeling 
lengths than for the positrons. 
The penetration lengths $L_{\mathrm p1}$ and 
$L_{\mathrm p1}$ for the electrons appear to be close to each other within the statistical 
uncertainties, for almost all the bending profiles studied in 
Table~\ref{Table_ep-lamu-fixed}, as well as to be much shorter than the thickness 
$L=150$~micron for the simulated crystalline sample. 
Therefore, either of these lengths can be regarded as the de-channeling length for the electrons. 
For the positrons, two penetration scales, $L_{\mathrm p1}$ and $L_{\mathrm p2}$, appear 
to be different, with the values $L_{\mathrm p2}$ being systematically smaller than 
the values $L_{\mathrm p1}$. 
For small bending amplitudes, the positron penetration scales are not much shorter than the 
thickness of the crystals studied, and therefore non of these scales are close to the 
de-channeling lengths. 
Estimating the positron de-channeling length on the grounds of the continuous potential 
approximation (Eq.~(1.50) of Ref.~\cite{BiryukovChesnokovKotovBook}) yields for the straight Si(110) 
crystal the length $\Ld \approx 570$~micron essentially longer than the crystalline 
thickness. 
Therefore, the values for the positron penetration scales $L_{\rm p1}$ presented in 
Table \ref{Table_ep-lamu-fixed} can be considered only as 
the lower bounds for the positron de-channeling length.

\begin{figure} [ht]
\centering
\includegraphics[scale=0.45,clip]{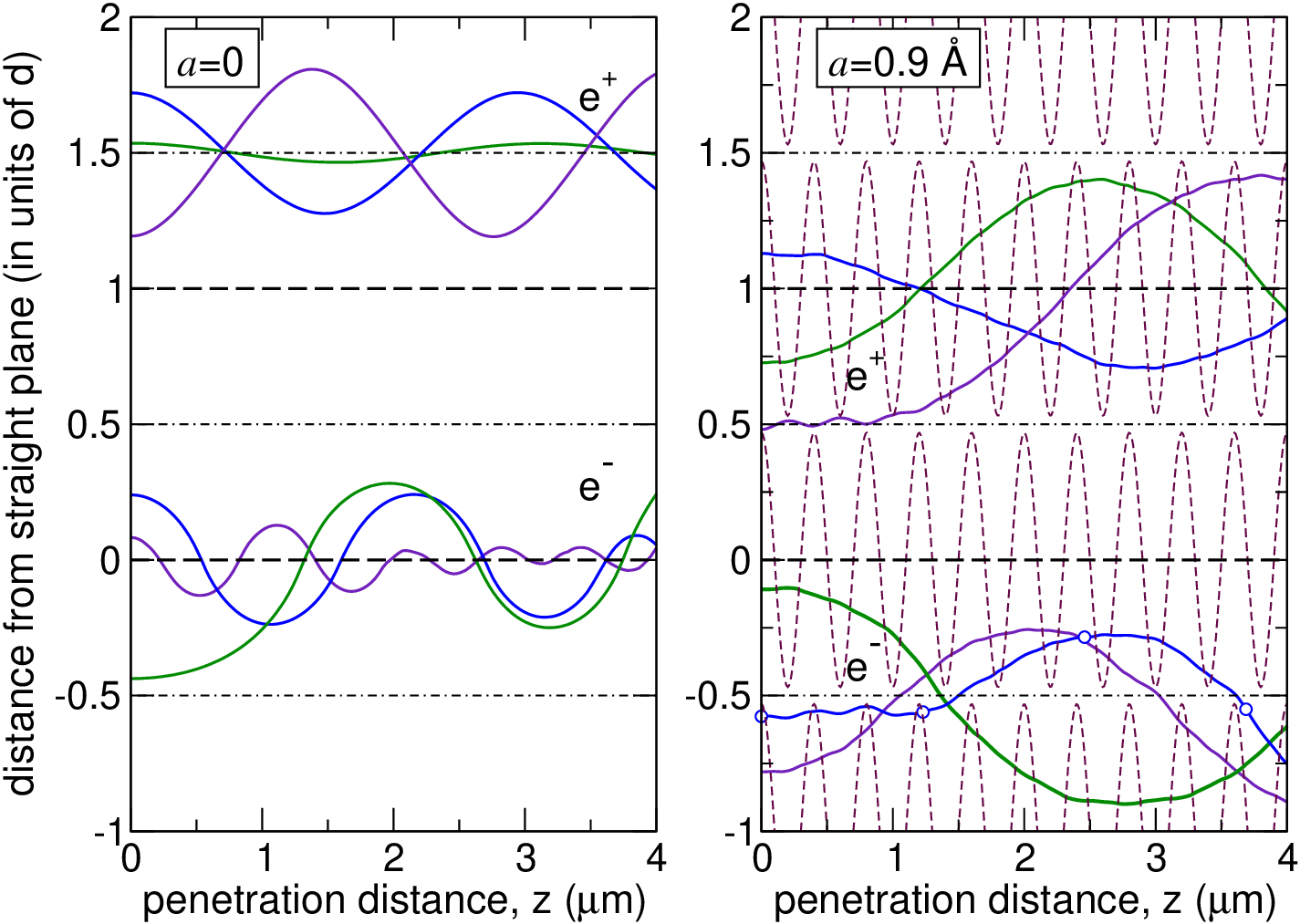}
\caption{
Channeling trajectories for the electrons($e^{-}$) and positrons 
($e^{+}$) in the straight (left plot) and periodically bent (right plot) Si(110). 
Thick dashed lines correspond to the straight (110)-planes, chained lines mark 
the centerlines between the planes. Thin dashed lines in the right plot show the 
profiles of the planes bent with $a=0.9$ \AA{} and $\lamu=400$ nm. The trajectories 
shown in the left and right plot can be regarded as ``conventional'' and 
``complementary'', respectively. 
Notice the shift $d/2$ ($d$ is the inter-planar distance) between the 
equilibrium positions for the transverse oscillations of the projectiles 
in the left- and right-plot trajectories.
}
\label{fig.01} 
\end{figure}

A close look at the trajectories simulated for various bending amplitudes reveals the 
channeling segments in the trajectories to be of two different kinds that can be called 
``conventional'' and ``complementary''. 
The fraction of ``complementary'' segments in the 
simulated trajectories is small and the corresponding channeling lengths are very short 
for the bending amplitudes $a \leq 0.2$ \AA{}. 
With increasing $a$ the latter fraction 
extends and dominates over the ``conventional'' fraction as the amplitude approaches the 
value $d/2$ with $d$ being the inter-planar distance in straight Si(110). 
The two kinds of the channeling trajectory segments yield essentially different 
values for the statistical quantities. 
In Table \ref{Table_ep-lamu-fixed}, the quantities 
deduced from the ``complementary'' segments are given in the lower lines for $a=0.7$, 
$0.8$ and $0.9$ \AA{}.
 All the other quantities in the table are computed from the 
``conventional'' segments. 

To explain the introduced terminology and to illustrate the dramatic change in 
the channeling processes for the projectiles with increasing $a$ we refer
to Fig. \ref{fig.01}. 
The left plot in the figure shows the channeling segments of the trajectories in the 
straight crystal. 
There, the projectiles exhibit the ``conventional'' channeling where the electrons oscillate 
around the crystalline planes whereas the positrons oscillate around
the centerlines between the planes. 
In the bent crystal studied in the right plot for $a=0.9$~\AA{} we encounter the 
``conventional'' channeling trajectory segments. 
There, the transverse oscillations appear around the equilibrium positions that 
are shifted in the inter-planar direction by the distance $d/2$ with respect to the 
equilibrium positions in the straight crystal. 
In other words, the equilibrium positions for the transverse oscillations of the electrons 
and the positrons are reversed in the ``conventional'' and ``complementary'' trajectory segments. 
The fraction of ``complementary'' segments in the simulated trajectories, 
being negligible for $a\leq 0.2$ \AA{}, increases with increasing bending amplitude 
and for $a\approx d/2$ almost all the channeling segments become ``complementary''. 

\begin{figure} [ht]
\centering
\includegraphics[scale=0.45,clip]{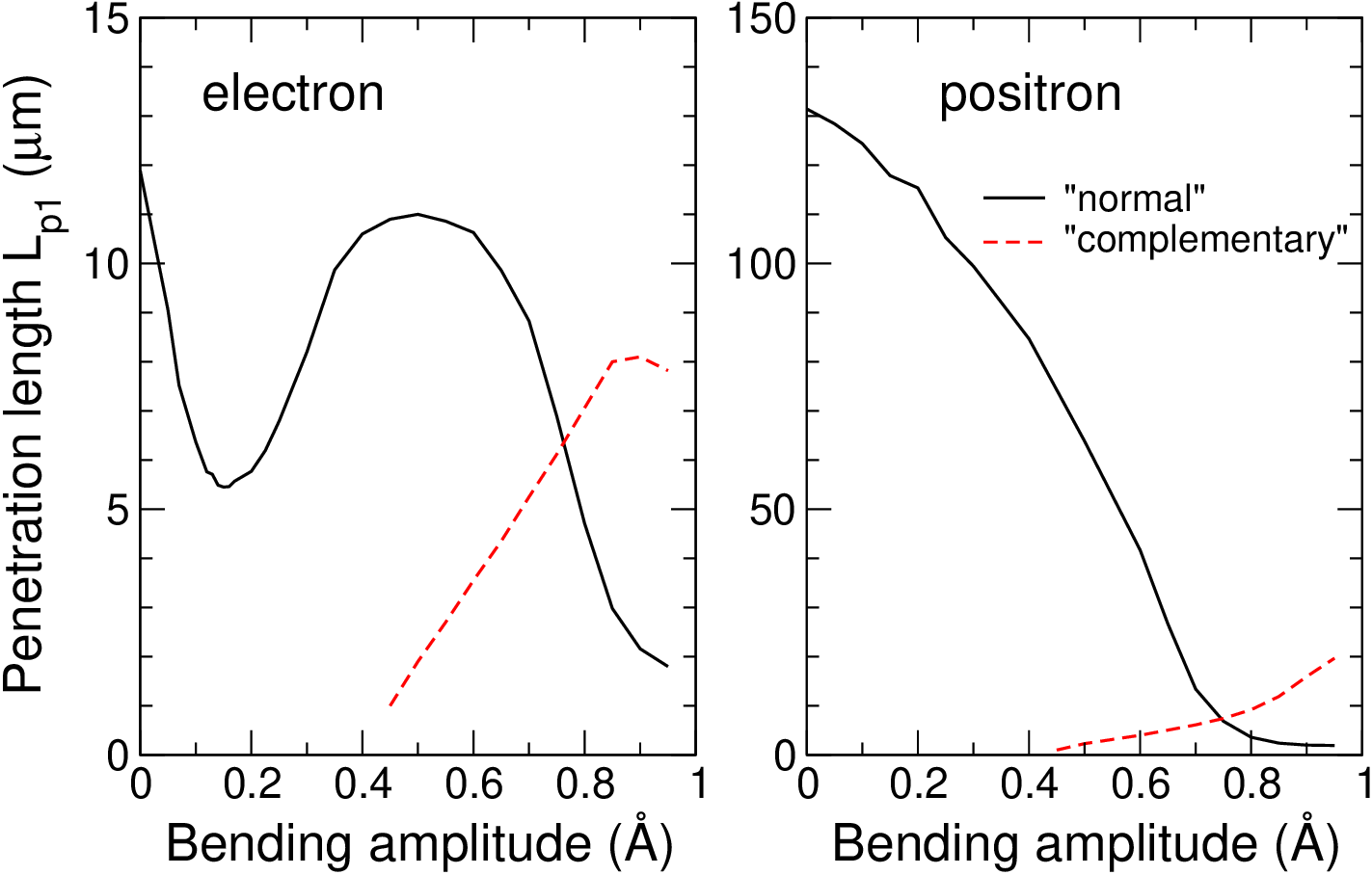}
\caption{
Penetration length  $L_{\rm p1}$ versus bending amplitude for the 855 MeV 
electrons (left plot) and positrons (right plot) in SASP bent Si(110). The bending 
period is fixed at $400$ nm. The solid and dashed curves represent the dependencies 
calculated from the ``conventional'' and ``complementary'' channeling trajectory 
segments, respectively. See also explanations in the text.
}
\label{fig.02} 
\end{figure}

The data presented in Table \ref{Table_ep-lamu-fixed} demonstrate that, the 
channeling lengths $L_{\rm p1,2}$ and $\Lch$ change with varying the bending 
period $a$ for the electrons and for the positrons in essentially different 
manner. To illustrate the difference, Fig. \ref{fig.02} shows the 
primary penetration lengths $L_{\rm p1}$ as functions of $a$ for the electrons 
(left plot) and the positrons (right plot). The length computed from 
the ``conventional'' trajectory segments (solid curves in the figure) 
vary monotonously with $a$ for the positrons and exhibits pronounced local 
minimum and maximum in the range $a\leq 0.6$ \AA{} for the electrons. 
The penetration lengths deduced from the ``complementary'' segments 
increase with increasing $a$ (dashed curves in the figure) 
being smaller than the ``conventional'' lengths unless $a$ approaches 
the values $\approx 0.75$ \AA{} for both types of the projectiles. 
For larger bending amplitudes the ``complementary'' channeling segments 
become on average longer than the ``conventional'' segments that 
display the values for the penetration lengths close to each other 
for the electrons and positrons. 

\begin{figure} [ht]
\centering
\includegraphics[scale=0.32,clip]{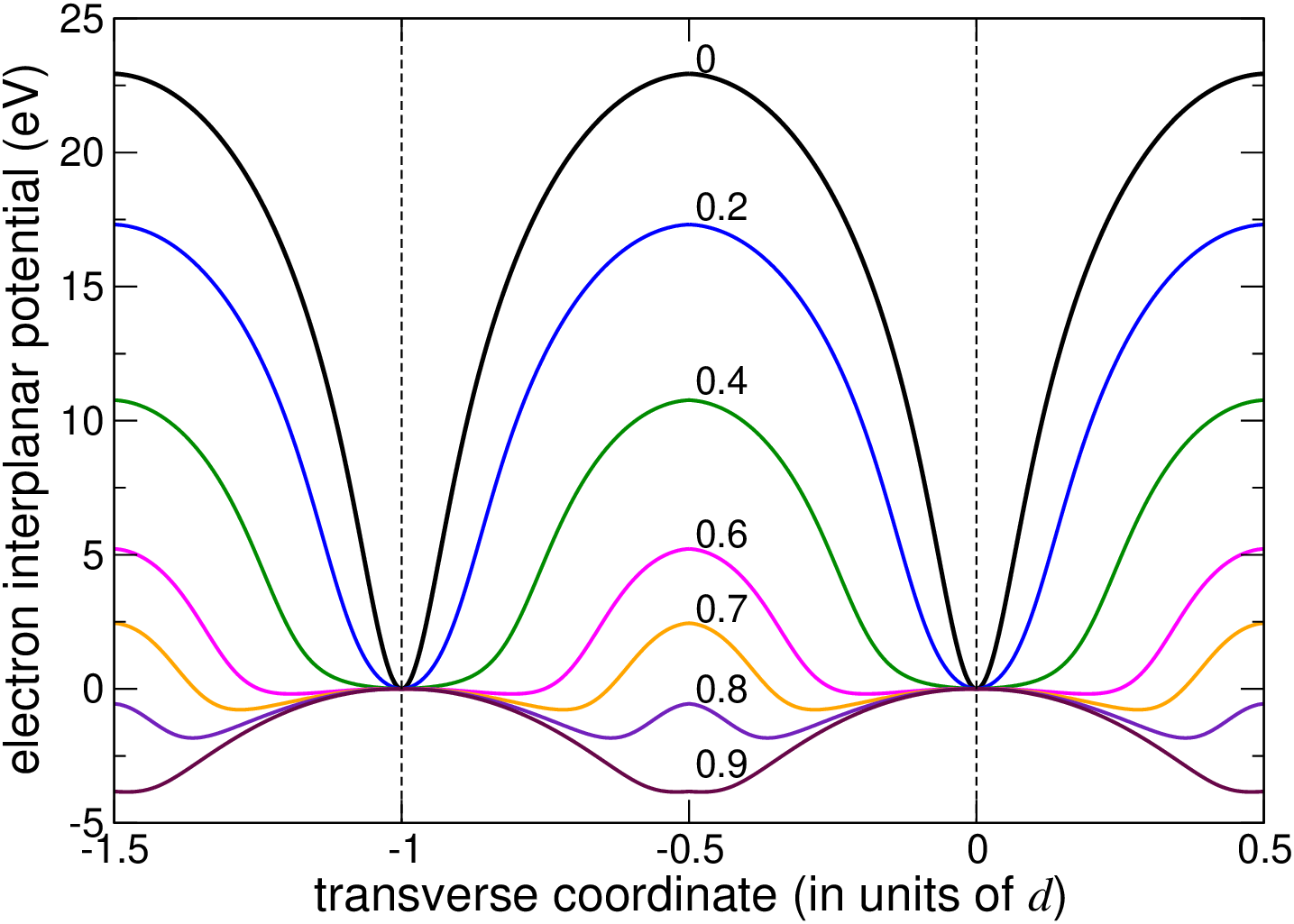}
\hspace*{0.3cm}
\includegraphics[scale=0.32,clip]{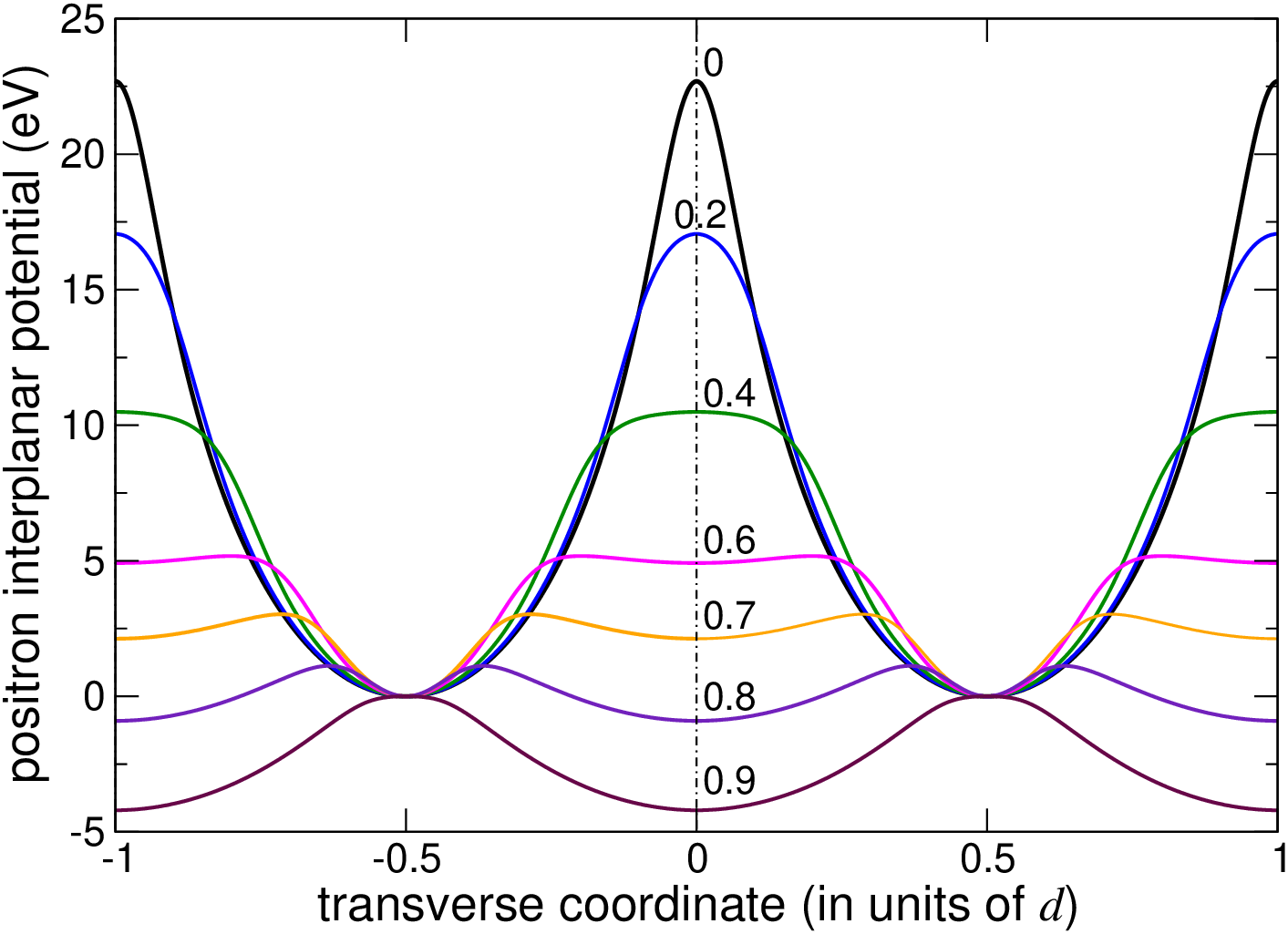}
\caption{
Continuous inter-planar potentials for the electrons (left plot) and 
positrons (right plot) for different values of the bending amplitude indicated 
in \AA{} near the curves ($a=0$ corresponds to the straight crystal). 
The potentials shown in the figure are evaluated for the temperature 
$300$~K by using the Moli\`{e}re atomic potentials and averaging the individual 
particle-atom interactions over the bending period $\lamu=400$~nm. 
The unit for the transverse coordinate in the plots is the inter-planar distance 
$d=1.92$~\AA{}. The vertical lines mark the adjacent (110)-planes in the 
straight crystal.
}
\label{fig.03}  
\end{figure}

\begin{figure} [ht]
\centering
\includegraphics[scale=0.32,clip]{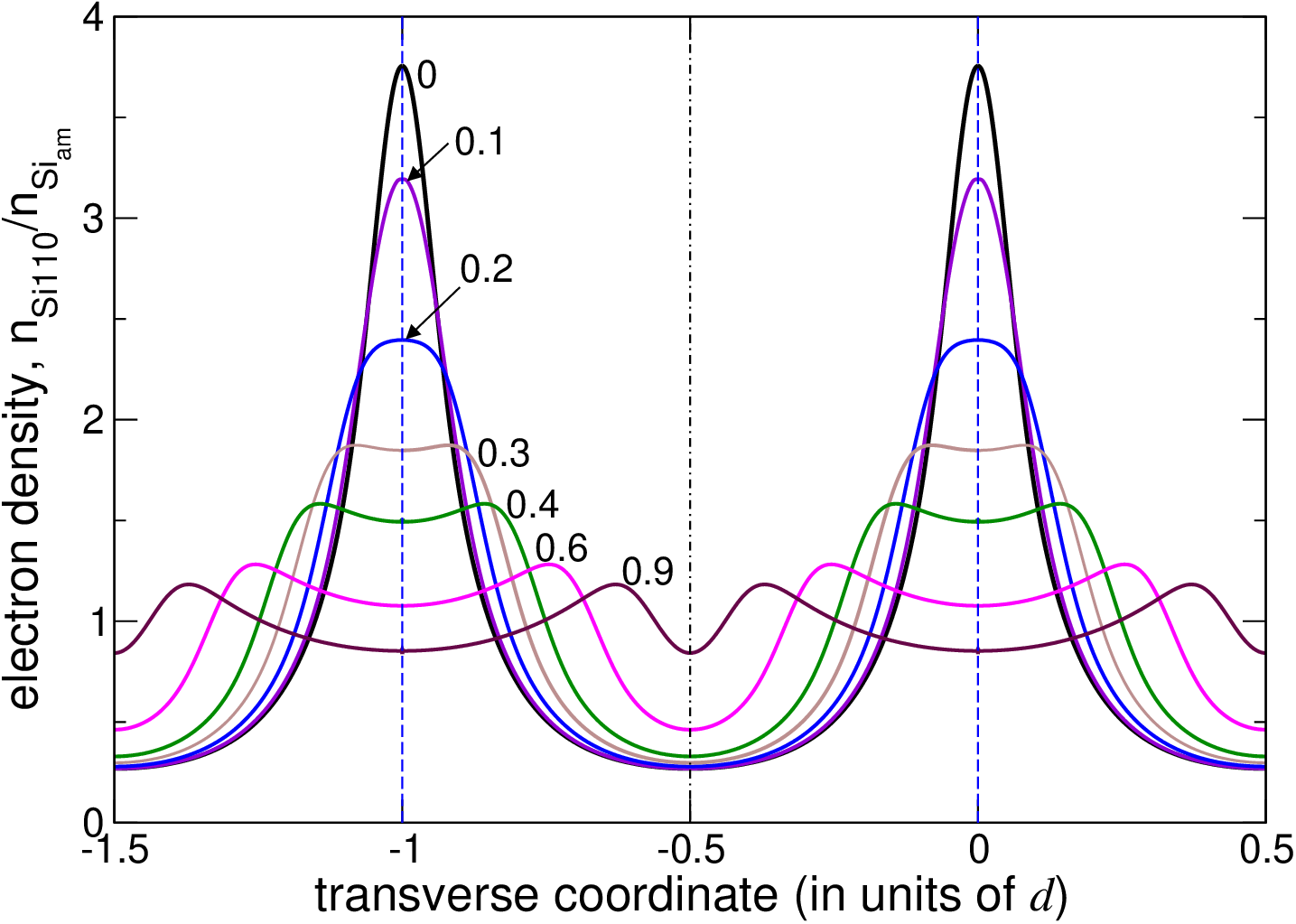}
\includegraphics[scale=0.32,clip]{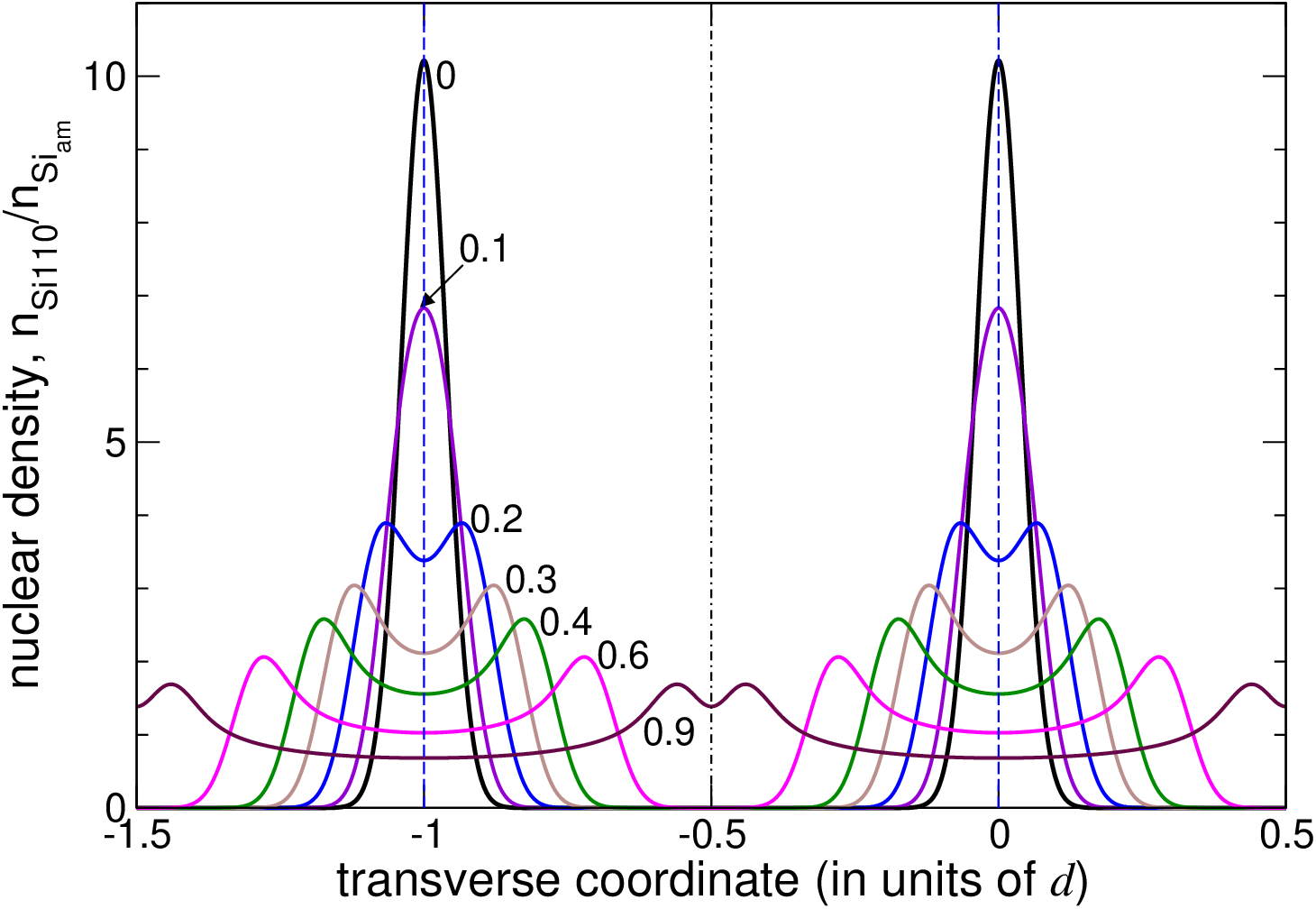}
\caption{
Distributions of the electronic (left graph) and nuclear (right graph) 
densities along the transverse coordinate. The densities are normalized with 
respect to the values for amorphous silicon and correspond to the potentials 
shown in Fig.~\ref{fig.03}. 
}
\label{fig.04}  
\end{figure}

The origin of the peculiarities described above are the changes in the particle-crystal 
interactions with changing the bending amplitudes. 
As already discussed, the positrons and electrons tend to channel in the domains with low
and high content of the crystalline constituents, respectively. 
In order to highlight how these domains change with changing the bending amplitude, 
a continuous inter-planar potential experienced by the moving projectiles can be considered. 
To calculate this potential for a periodically bent crystal we used the approach described 
in \ref{Pot:SASL-CU}. 
The non-periodic part of the emerging potential, Eq.~(\ref{Pot-CU:eq.10}), is 
evaluated by using the Moli\`{e}re atomic potentials. 
This part is presented in Fig. \ref{fig.03} for the electrons (left plot) 
and positrons (right plot), for different values of the bending amplitudes (given in \AA{} 
near the curves). 
Fig. \ref{fig.04} shows the corresponding 
volume densities of the crystalline electrons and nuclei. 

In Fig. \ref{fig.03}, the potential curves labeled with ``0'' 
correspond to the straight crystal. For small amplitude values, $a\leq 0.2$ \AA{}, 
the major changes in the potentials with increasing $a$ are decreasing potential 
barriers and, for the electrons, a broadening potential well. As the $a$ values 
approach $0.4\dots0.6$~\AA{}, the volume density of atoms becomes more friable and 
leads to flattening of the potential in the vicinity of the potential minimum for 
the electrons and of the potential maximum for the positrons. For larger amplitude 
values, the potentials change in a more dramatic way as the additional potential 
wells appear. These wells force 
the projectile to channel in spatial regions different from the channels for 
the straight and small-amplitude bent crystal. The latter regions are the 
``complementary'' channels that appear in vicinities of the centerlines for the 
periodically bent planes for the positrons, and are shifted away from the 
centerlines for the electrons. For the largest bending amplitude studied, 
$a=0.9$ \AA{}, the channeling can virtually be maintained only through the 
``complementary'' channels. Thus, the potentials and charge densities shown in 
Figs.~\ref{fig.03} and \ref{fig.04} 
elucidate evolution from ``conventional'' to ``complementary'' channeling of 
the projectiles with increasing the bending period.

The charge density distributions shown in Fig. \ref{fig.04} 
explain the peculiarities in the dependencies of the channeling lengths on the 
bending period. We reiterate that the densities are calculated within the same 
model as the continuous potentials studied in Fig. \ref{fig.03}. 
The vertical lines in Fig. \ref{fig.04} mark the centerlines 
of the ``conventional'' channels for the electrons (dashed lines) and the 
positrons (chained lines). It is seen that, with increasing $a$ both the 
electronic and nuclear crystalline densities gradually increase in the region 
between the two crystalline planes. As a result, motion of the positrons in 
the ``conventional'' channels is accompanied by increasing probability of 
collisions with the nuclei. This effect, together with lowering potential well, 
leads to a monotonous decrease of the channeling lengths. 
In contrast, in 
the ``complementary'' positron channels the densities decrease gradually, 
so that the corresponding channeling lengths increase. 
The non-monotonous variation with $a$ of the lengths for ``conventional'' 
channeling of the electrons can be interpreted as follows. 
For small values 
of $a$, the decrease in the potential barrier out-powers the decrease in the 
densities in the central part of the channel. 
As a result, the channeling lengths 
initially decrease with increasing $a$. For $0.2 \leq a \lesssim 0.5\dots 0.6$ \AA{}, 
the density distributions become more diffuse with noticeably lower magnitudes 
in the center of the channel, making thereby the de-channeling process to develop 
more effectively. 
However, the inter-planar potential barrier still remains 
sufficiently high and supports the growth of $L_{\rm p1,2}$ and $\Lch$ with $a$. 
For the amplitudes larger than $0.6$~\AA{}, the densities become comparable with 
those in the amorphous medium. Together with gradually decreasing potential 
barrier this yields decreasing channeling lengths.

Similar arguments referring to the continuous inter-planar potentials and charge densities, 
also explain the behavior of the electron and positron acceptances with varying bending 
amplitude. These argument readily confirm the particular outcome of the simulations that, 
for large amplitude values, $a\geq 0.6$ \AA{}, the acceptances by the ``complementary'' 
channels become comparable or even larger than the acceptances by the ``conventional'' 
channels. However, it remains to elucidate why the sum of the two acceptances can exceed 
the unity value, which might confuse one having in mind the meaning of acceptance as a 
{\em fraction} of the incident particles captured into the channeling mode at the 
entrance. The artifact values for the sum of ``conventional'' and ``complementary'' 
acceptances arise from double-counting of some particles as being accepted by both 
kinds of channels. Channeling in a SASP bent crystal develops with the trajectory 
oscillations of distinctly different periods, the long period $\lamch$ of channeling 
oscillations and the short period $\lamu \ll \lamch$ of oscillations due to the bending. 
At the entrance, a particle experiencing the short-period oscillations is distinguished 
in the simulations as channeling regardless of the type, ``conventional'' or 
``complementary''. For illustration, we refer to the electron trajectory marked with the 
open circles in the right plot of Fig. \ref{fig.01}. Near the crystalline 
entrance, this trajectory exhibits several oscillations with the period $\lamu$ becoming 
thereby accepted by both the ``conventional'', $-1.5 < y/d <-0.5$, and the 
``complementary'', $-1 < y/d < 0$, electron channels.

\subsection{Radiation Spectra \label{Spectra}}

\begin{figure} [ht]
\centering
\includegraphics[scale=0.32,clip]{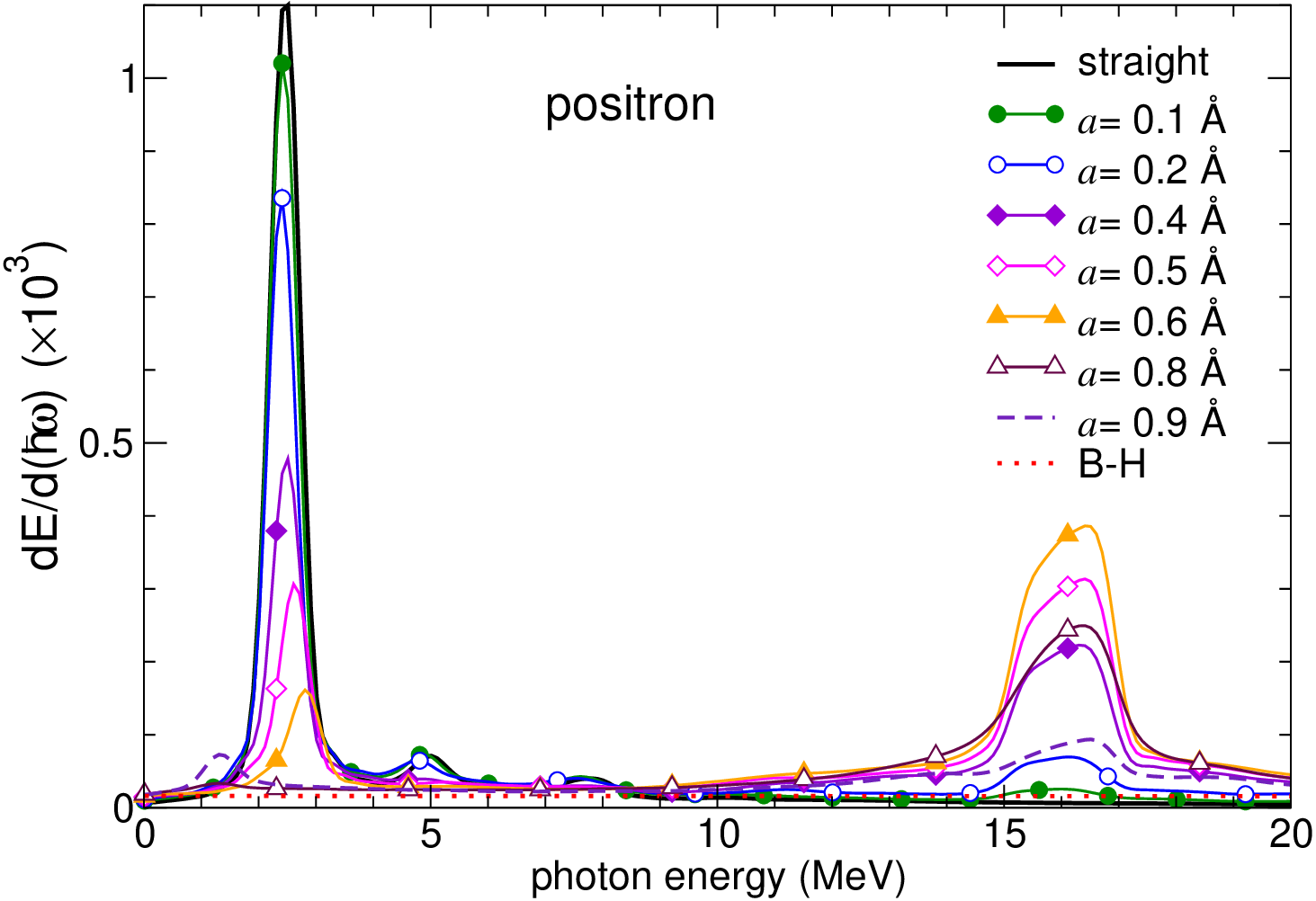}
\includegraphics[scale=0.32,clip]{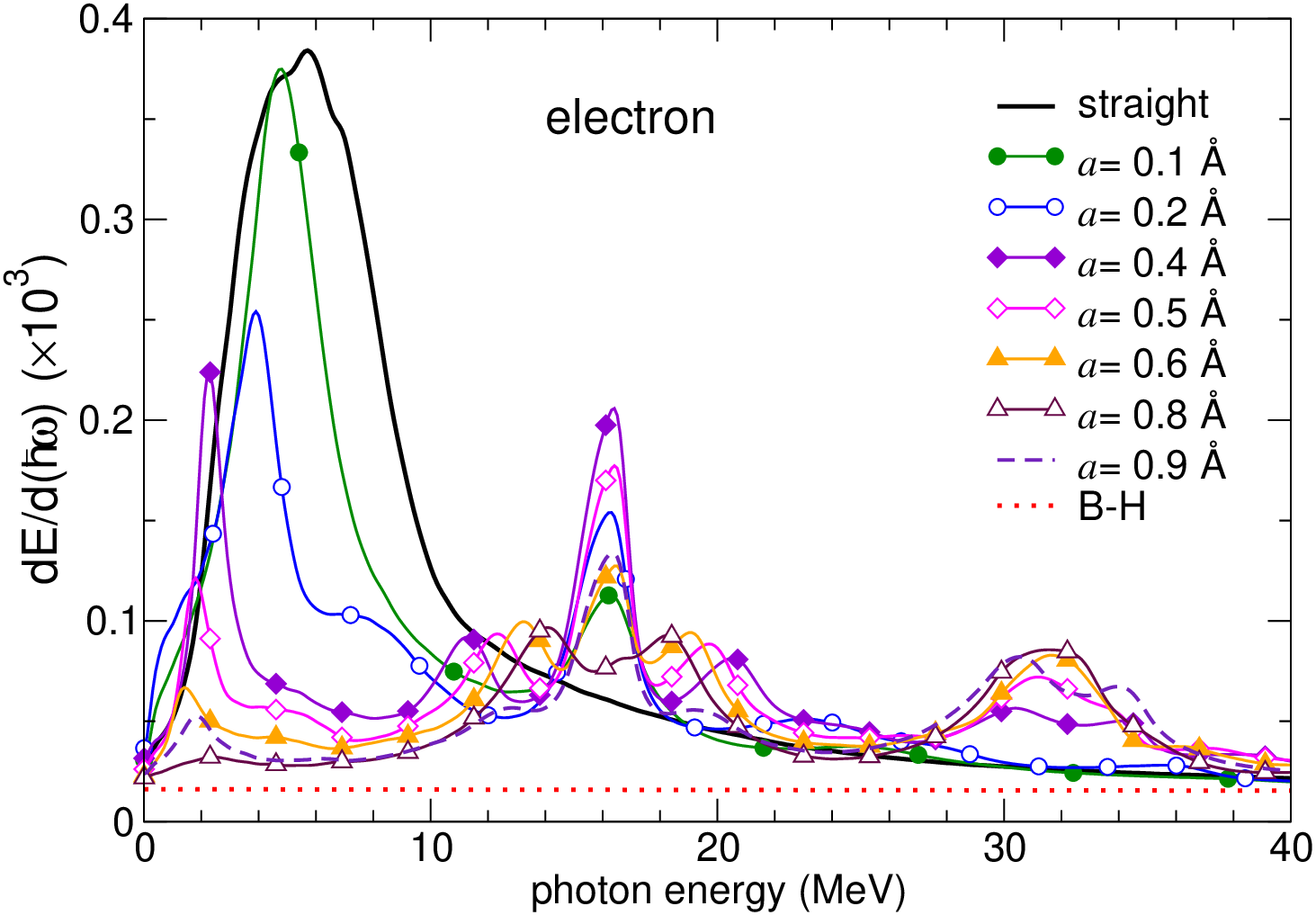}
\\
\vspace*{0.3cm}
\includegraphics[scale=0.32,clip]{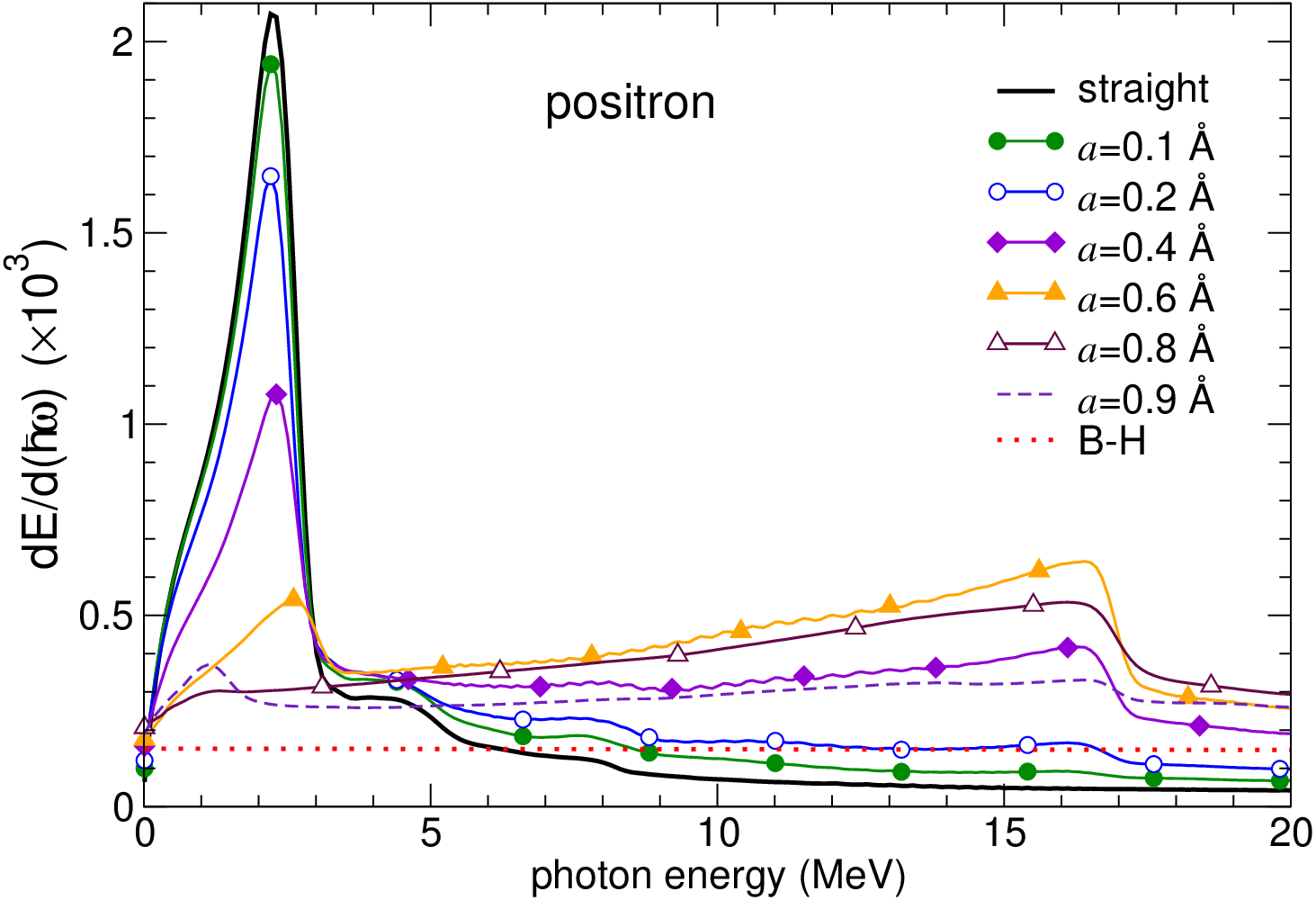}
\includegraphics[scale=0.32,clip]{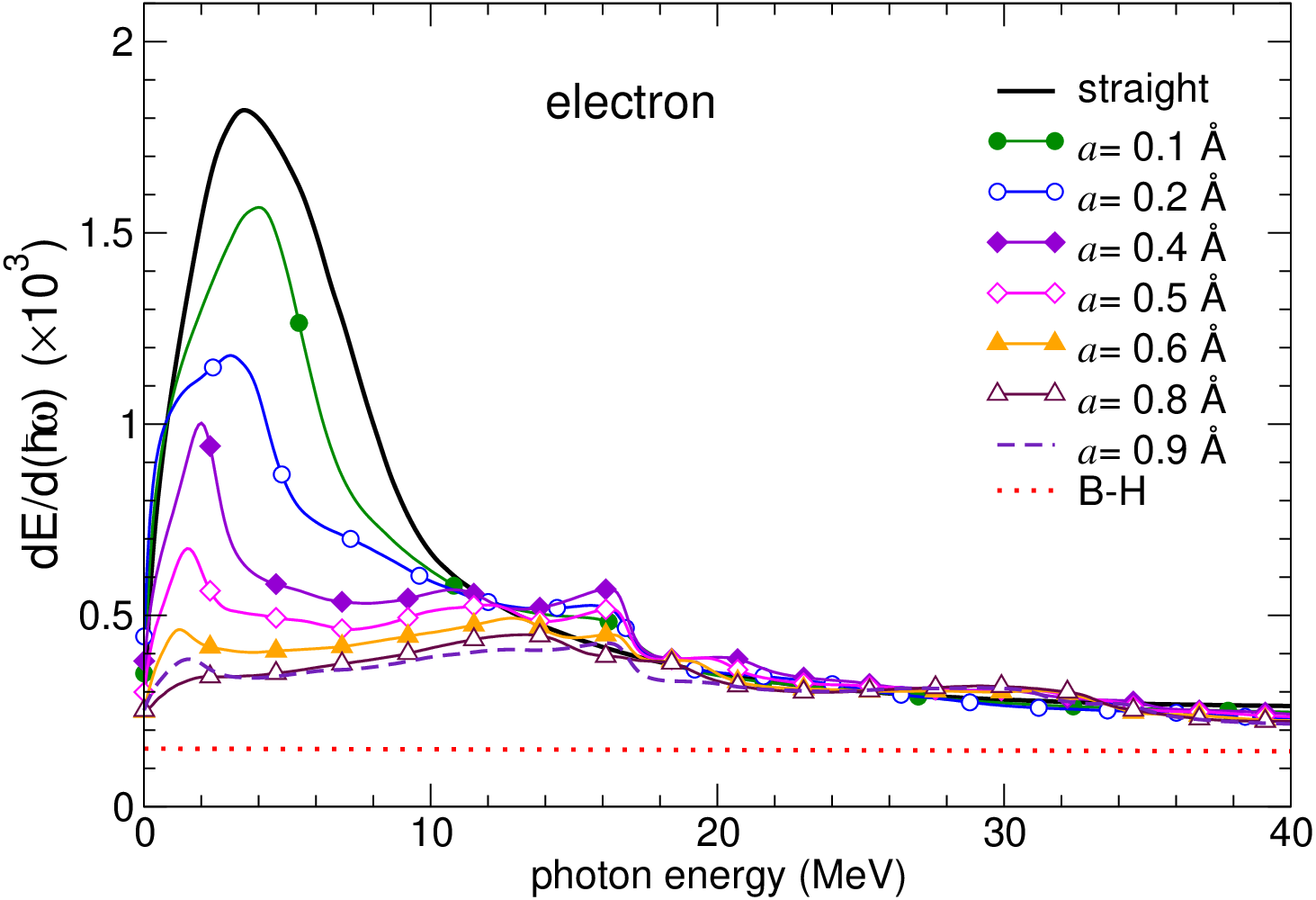}
\caption{
Spectral distribution of radiation emitted by 855 MeV positrons (left) 
and electrons (right) in straight (thick solid lines) and SASP bent Si(110) 
with the period $\lamu=400$ nm and various amplitudes $a$ as indicated. 
Dotted lines mark the Bethe-Heitler spectra for amorphous silicon. 
The upper and lower plots correspond to the aperture values 
$\theta_{\max} = 0.21$ mrad and $\theta_{\max} = 4$ mrad, respectively. 
All spectra refer to the crystal thickness $L=12$ micron.}
\label{fig.05}
\end{figure}

\begin{figure} [ht]
\centering
\includegraphics[scale=0.32,clip]{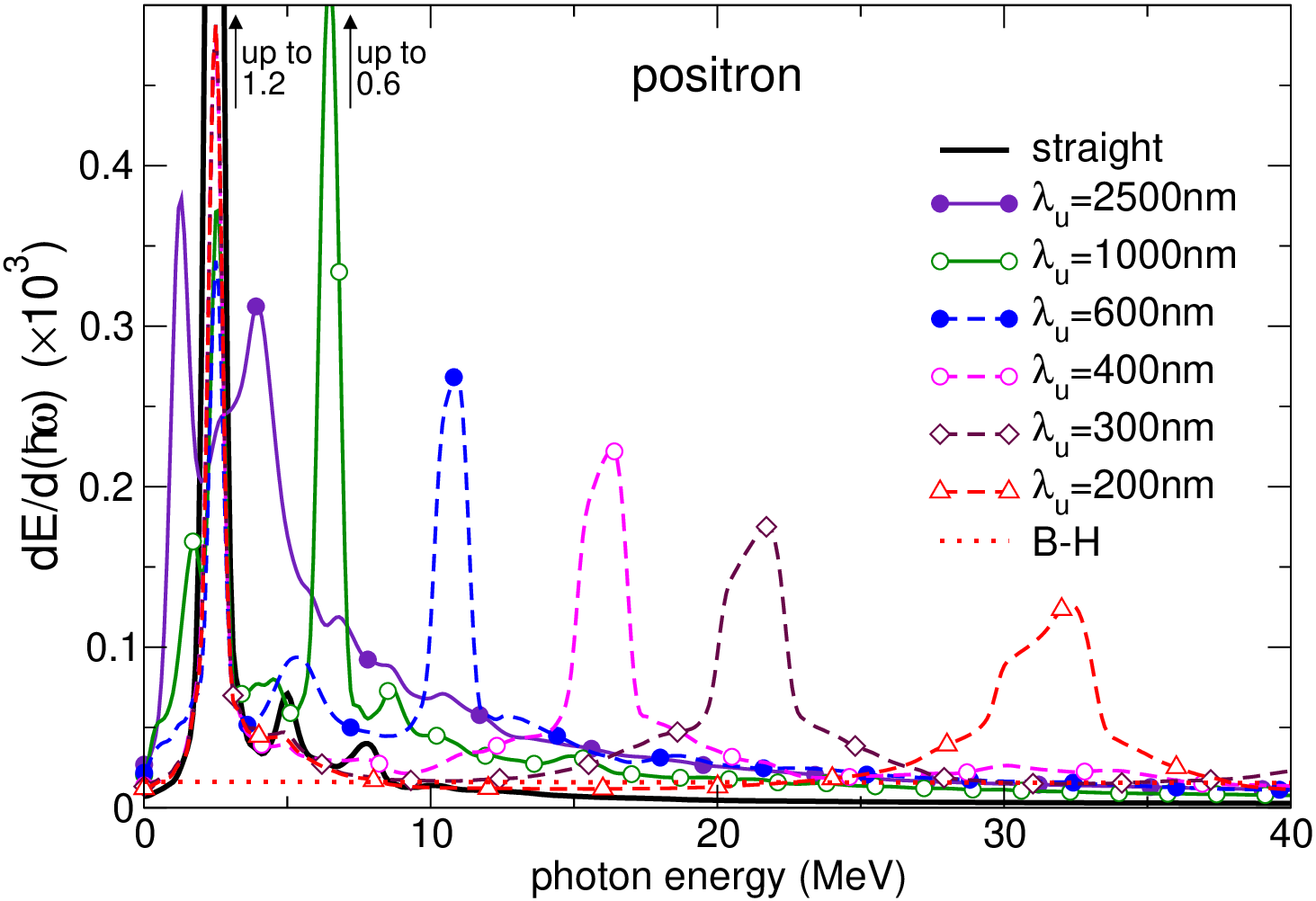}
\includegraphics[scale=0.32,clip]{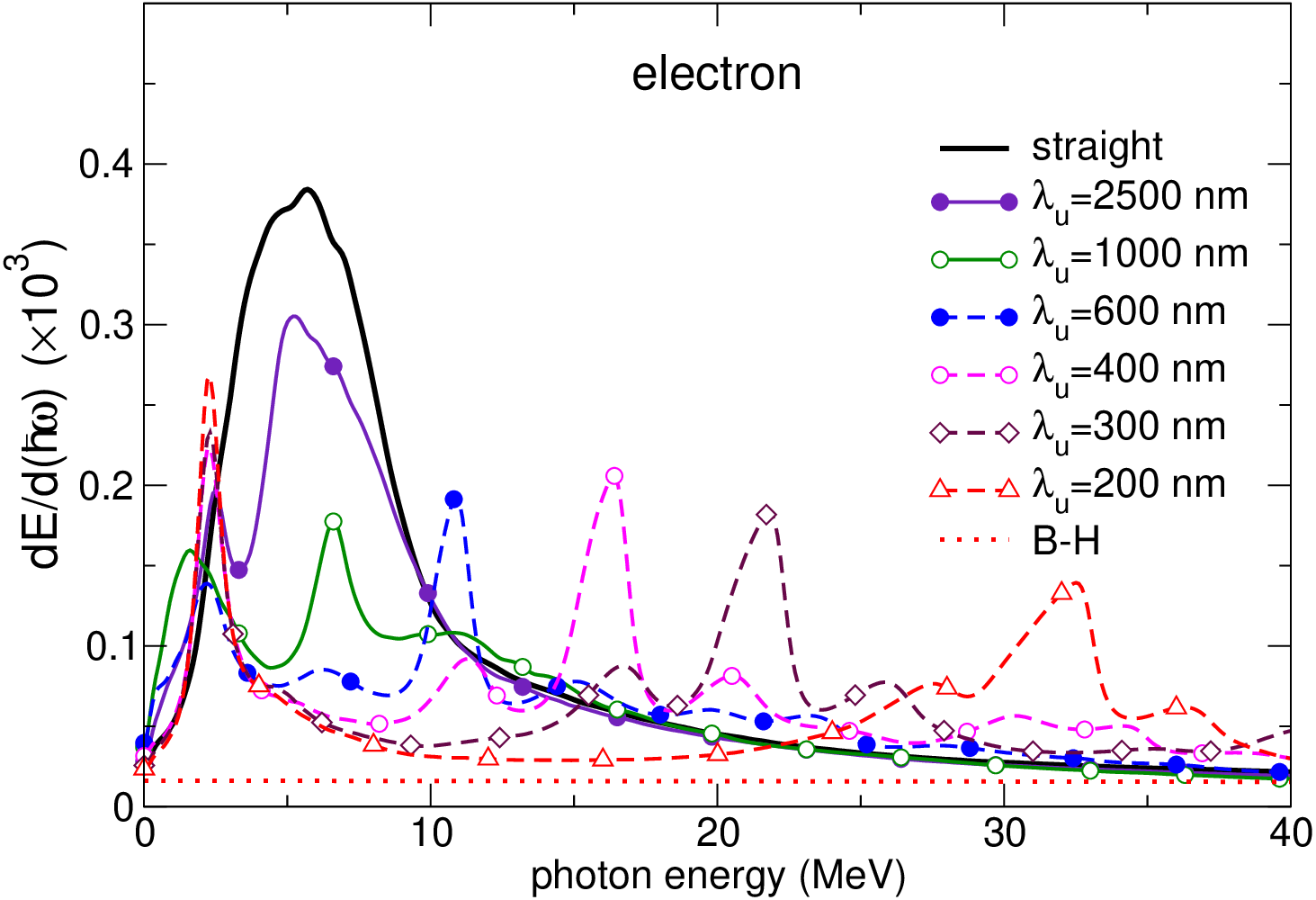}
\\
\vspace*{0.3cm}
\includegraphics[scale=0.32,clip]{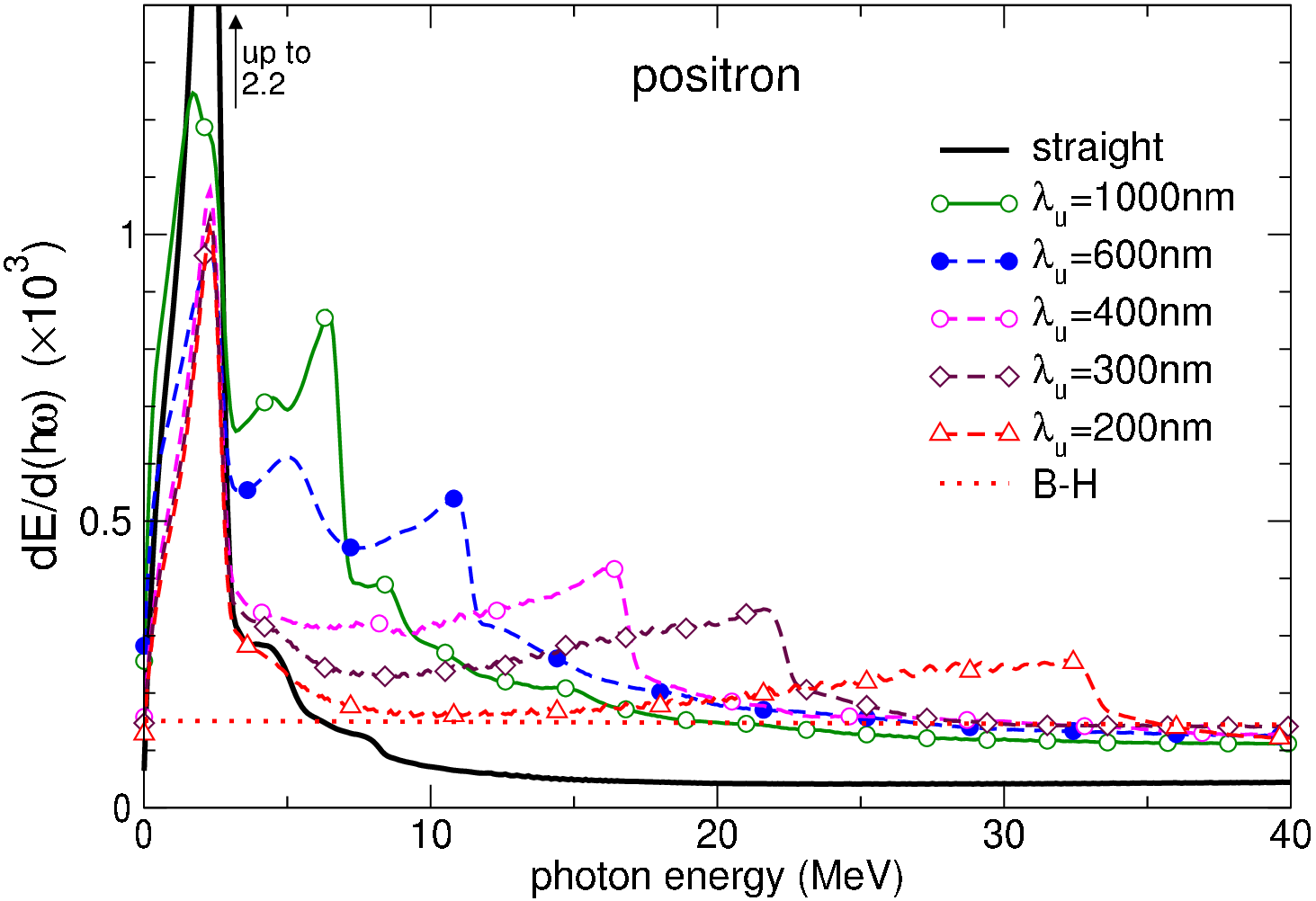}
\includegraphics[scale=0.32,clip]{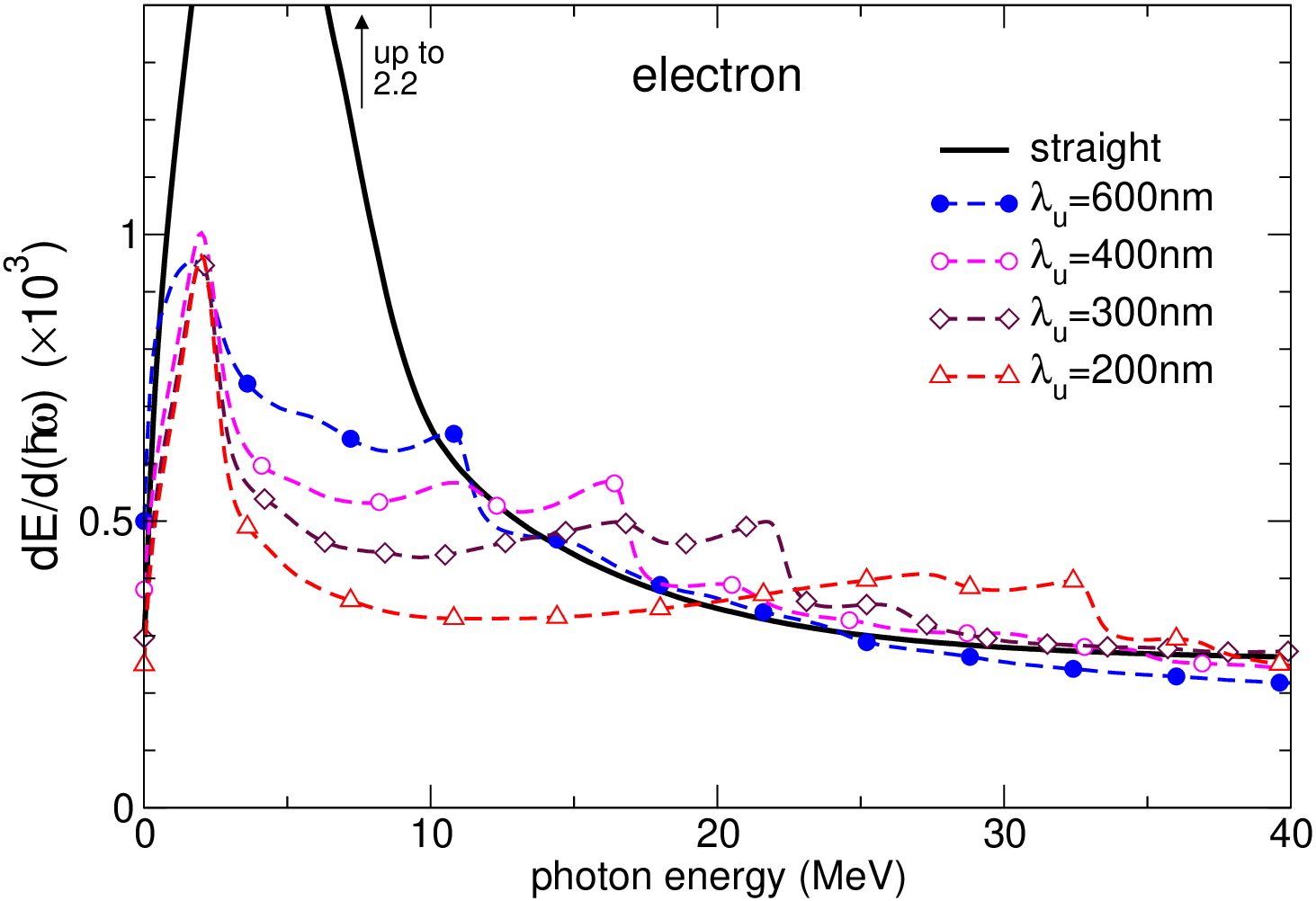}
\caption{
Spectral distribution of radiation emitted by 855 MeV positrons (left) and electrons 
(right) in straight (thick solid curves) and SASP bent Si(110) with the 
bending amplitude $a=0.4$ \AA{} and various periods $\lamu$ as indicated. 
Dotted lines mark the Bethe-Heitler spectra for amorphous silicon. 
The spectra are computed for the crystal thickness $L=12~\mu$m, and two values for 
the radiation aperture, $\theta_{\max} = 0.21$ mrad (upper plots) and 
$\theta_{\max} = 4$ mrad (lower plots). 
} 
\label{fig.06}
\end{figure}

The above described statistical studies on the channeling properties show 
the penetration lengths for the $855$~MeV electrons to not exceed a value of 
about $12~\mu$m. We have therefore opted to study first the radiation for the 
$12~\mu$m thick silicon crystal. The radiation spectra produced by the 
particles incoming along the (110) crystallographic planes are presented 
in Figs. \ref{fig.05} and \ref{fig.06}. 
In Fig. \ref{fig.05}, the spectra are studied for 
the fixed value of the bending period, $\lamu = 400$ nm, and different 
values of the bending amplitudes in the range $a=0\dots0.9$ \AA{}. 
Fig. \ref{fig.06} presents the spectra produced 
for SASP bending with the amplitude $a=0.4$ \AA{} and various periods 
$\lamu = 200\dots 2500$ nm. The sets of plots in both figures relate to 
the different projectiles, electrons and positrons 
(right and left plots, respectively), and different values for the 
radiation aperture, $\theta_{\max}=0.21\ \mbox{mrad}$ and 
$\theta_{\max} = 4$~mrad (top and bottom, respectively). 
The smaller aperture value refers to a nearly forward 
emission from $855$~MeV projectiles ($\theta_{\max} \approx (3\gamma)^{-1}$) 
wheres the second value corresponds to the emission cone collecting almost all 
the radiation from the relativistic particles ($\theta_{\max} \gg \gamma^{-1}$). 

The spectra computed for the straight crystal and various bending amplitudes display a 
variety of features seen in Fig. \ref{fig.05}. 
To be noticed are the pronounced peaks of the channeling radiation in the spectra for 
the straight crystal (the black solid-line curves). Nearly perfectly harmonic 
channeling oscillations in the positron trajectories (the examples of the simulated 
trajectories can be found in \cite{ChanModuleMBN_2013,Sub_GeV_2013,ChannelingBook2014}) 
lead to the undulator-type spectra of radiation with small values of the undulator 
parameter, $K<1$ (see, e.g., \cite{Baier}). The radiation spectra produced by the 
positrons in straight Si(110) clearly display the fundamental peaks of the channeling 
radiation at the emission energy $\approx 2.5$~MeV, whereas the higher harmonics 
appear to be strongly suppressed according to the small values of $K$. 
In particular, for the smaller aperture value, $\theta_{\max} = 0.21$~mrad, the 
maximal spectral spectral intensity in the fundamental peak is an order of magnitude 
larger than that for the second harmonics displayed by a small peak at about $5$~MeV, 
and only a tiny hump of the third harmonics of channeling radiation can be recognized 
at about 7.5~MeV (see the top left plot in the figure). 
For the electrons passing through the straight crystal, the channeling radiation peaks 
are less intensive and much broader than these for the positrons, as a result of 
stronger anharmonicity of the channeling oscillations in the trajectories. 
For the larger aperture value, $\theta_{\max} = 4$~mrad, a sizable part of the 
energy is radiated at the angles $\gamma^{-1} < \theta < \theta_{\max}$. 
For relatively large emission angles, the emission energy for the first channeling 
spectral harmonics decreases with increasing angle. As a result, the fundamental peaks 
of channeling radiation broaden and shift towards softer radiation energies. 

The radiation spectra produced by the projectiles passing the SASP bent crystals 
display additional peaks, more pronounced for the smaller aperture value, which 
emerge from the short-period modulations of the projectile trajectories by 
the bent crystalline structure. The major novel feature of the radiation is that 
the peaks due the bending appear at the emission energies larger than the energies 
of the channeling peaks. For both types of the projectiles, the fundamental spectral 
peaks in the radiation emergent from SASP bending correspond to the emission energy 
about $16$~MeV significantly exceeding the energy $2.5$~MeV for the fundamental 
channeling spectral peaks. For the positrons, the peaks of radiation due to the 
bending are displayed in the spectra for the amplitude values 0.3\dots0.8 \AA{}. 
For smaller values of $a$, the spectral peaks disappear because the positrons 
experience mainly ``conventional'' channeling staying away from the crystalline 
atoms and being therefore less affected by the SASP bent planes. In contrast, 
the electrons experience the impact of the bent crystalline structure at lower 
bending amplitudes. As seen in the right upper plot for the fundamental spectral 
peaks emergent from the bending, the peak for $a=0.1$ \AA{} is only two times 
lower than the maximal peak displayed for $a=0.4$ \AA{}.

To be noted are the spectral properties displayed for larger aperture value 
(bottom plots in Fig.~\ref{fig.05}). In the spectra produced by 
the electrons for the values of $a$ exceeding $0.2$ \AA{}, noticeable are the lines at the 
energies around 32 MeV which are clearly the second harmonics of the radiation emergent from 
the SASP bending. The impact of increasing bending amplitude on the lines of channeling 
radiation for the electrons are decreasing line heights accompanied with line shifts towards 
the lower emission energies. In contrast, the positron spectra exhibit less peculiarities 
gradually converging to the Bethe-Heitler background with increasing radiation energies. 
Above $20$~MeV, the radiation spectra for the positrons become fairly close to the 
background spectra and are not shown in the figure.

The impact of different periods of SASP bending on the radiation spectra is studied in 
Fig.~\ref{fig.06}. The positions of the lines due to the bending 
are clearly seen to be in reciprocal relation with the bending period and shift towards 
hard-range radiation energies with decreasing $\lamu$. In contrast, the positions of 
channeling spectral lines, especially for the positrons, do not noticeably change 
with varying $\lamu$. Yet the shapes of the channeling lines are different 
for different bending amplitudes: the lines decrease in height as well as slightly 
shift towards the softer radiation energies with increasing bending period. The changes 
in the channeling radiation develop when the channeling lines and the lines resulting from 
the bending appear close to each other in the spectra and therefore interfere. The latter 
effects are more prominent for the larger aperture value (bottom plots in the figure), 
and in the spectra for the electrons (right plots) then in the spectra for the positrons 
(left plots).

\begin{figure} [ht]
\centering
\includegraphics[scale=0.32,clip]{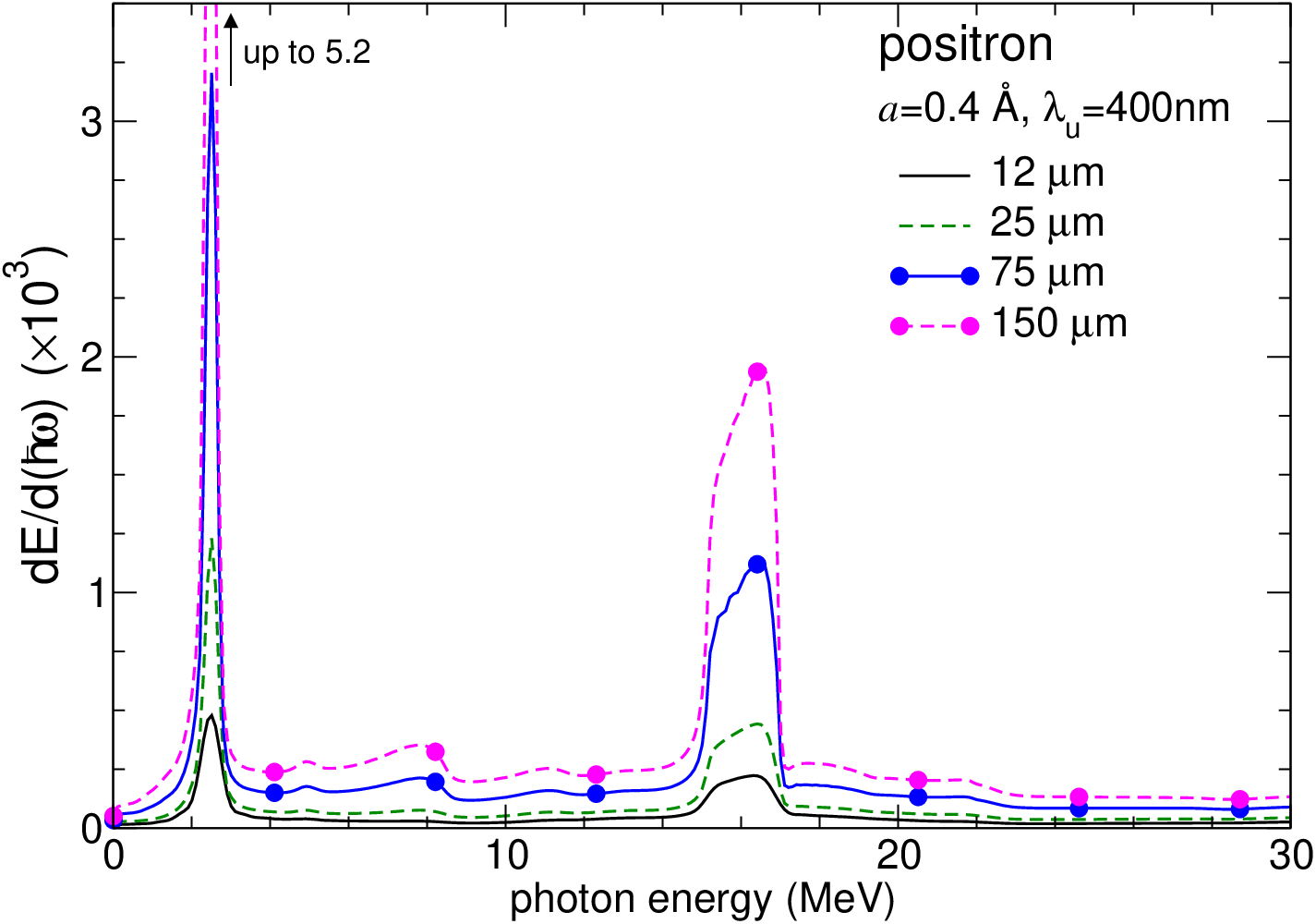}
\includegraphics[scale=0.32,clip]{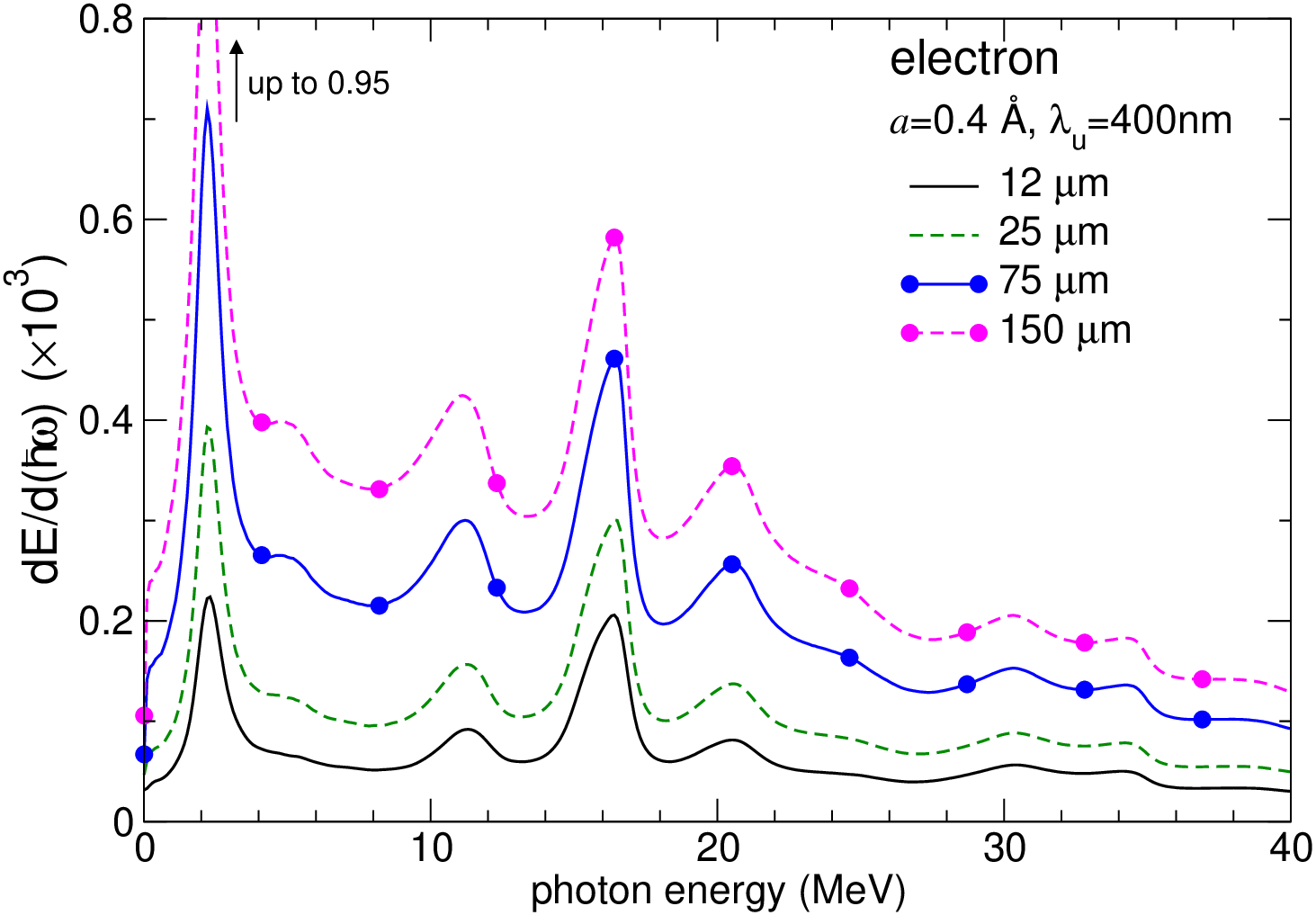}
\\
\vspace*{0.3cm}
\includegraphics[scale=0.32,clip]{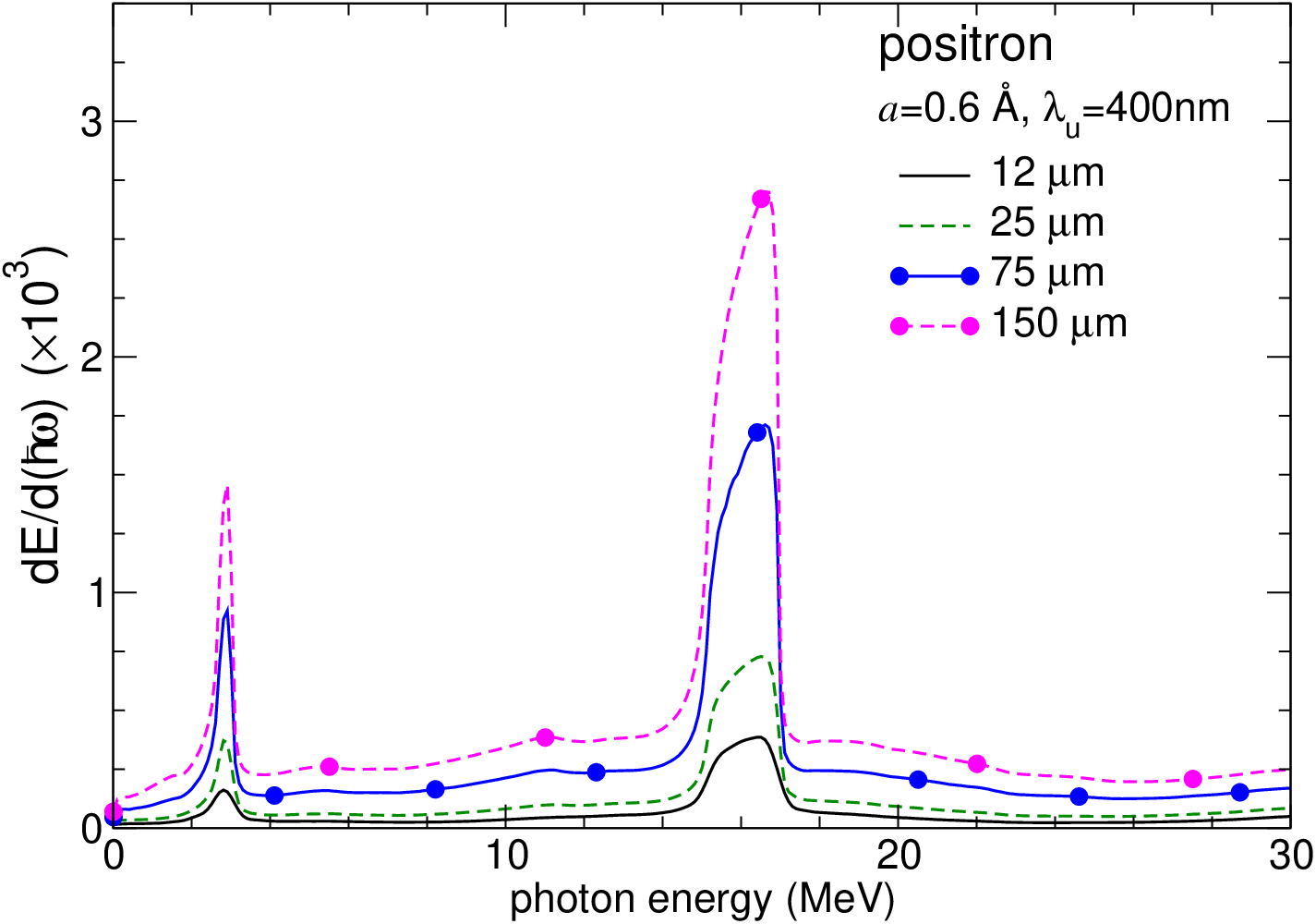}
\includegraphics[scale=0.32,clip]{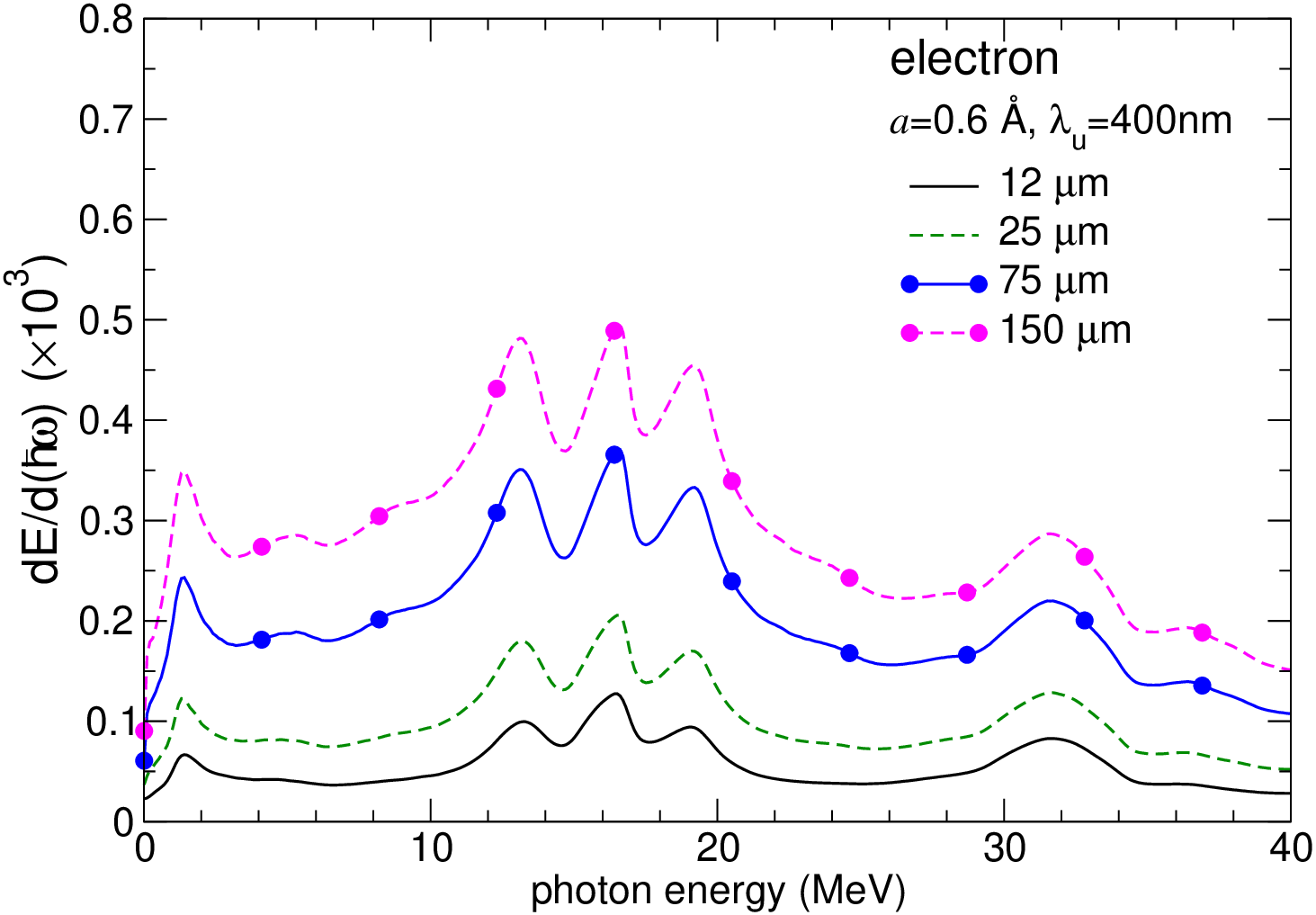}
\caption{
Spectral distributions of radiation emitted by the $855$~MeV electrons (right plots) 
and positrons (left plots) for SASP bending with the period $\lamu=400$~nm and 
the amplitudes $a=0.4$~\AA{} (upper plots) and $0.6$~\AA{} (lower plots). 
The spectra are computed for the radiation aperture $\theta_{\max} = 0.21$~mrad and 
different crystal thicknesses indicated in the plots. 
}
\label{fig.07}
\end{figure}

To complete the studies on the radiation from SASP bent Si(110), we show in 
Fig.~\ref{fig.07} the spectra simulated for four different thicknesses 
of the crystal, from $12$ up to $150$ micron. We can conclude that, the spectral shapes remain 
fairly the same with increasing crystalline thickness, with naturally increasing spectral 
intensities. The spectra in the figure also display the effects of varying bending amplitude. 
For both types of the projectiles, increasing the amplitude from $0.4$~\AA{} to $0.6$~\AA{} 
results in suppressing the channeling spectral lines and amplifying the lines produced due 
to the bending. 

\section{Conclusions \label{Conclusions}}

In this paper, we have provided a systematic analysis of the channeling and radiation 
in short-period small-amplitude bent silicon crystals. 

The statistical properties of channeling were described in terms of the lengths quantifying 
the spatial scales of de-channeling and re-channeling processes experienced by the projectiles. 
We have particularly focused on behavior of the lengths with varying the bending amplitude 
of the crystalline planes. With the amplitude increasing above already some moderate values, 
we have encountered a drastic change in the channeling process for the both types of 
projectiles. To elucidate the underlying physics, an analytical model has been developed 
in terms of the charge densities for the bent crystalline media which influence the 
particles moving through the crystal. The model reveals the channels optimal for ``binding'' 
the transverse motion of the projectiles to shift in the inter-planar direction with increasing 
the bending amplitude. The varying properties of channeling can be clarified in terms of 
two groups of the trajectories, the ``conventional'' and ``complementary'' ones.  
The supplementary analytical model of the continuous potential turns out to be helpful 
proving the simulations performed with \textsc{MBN Explorer} to be reliable, in particular, 
with respect to advance account for the interaction of the projectiles with individual atoms 
of the crystalline media. 

\section*{Acknowledgments}

The work was supported by the European Commission (the PEARL Project
within the H2020-MSCA-RISE-2015 call, GA 690991).

\appendix

\section{Continuous Potential in a SASP Bent Crystal 
\label{Pot:SASL-CU}}

In this supplementary section we develop an approximation of continuous inter-planar 
potential. This potential and the corresponding distributions of the crystalline 
charge densities help in qualitative explanations of the results of numerical 
simulations on the motion and radiation for the electrons and positrons in SASP 
bent crystal, discussed above in Sects. \ref{Lengths} and \ref{Spectra}.      

\subsection{Continuous Potential of a Periodically Bent Plane \label{Pot:CU}}

To derive the approximation of continuous potential in a crystal with periodically 
bent planes we use a general approach developed to study the radiation by fast 
projectiles in acoustically excited crystals (see, for example, Ref.~\cite{Dedkov1994}).

For the cosine periodic bending (\ref{PB_profile:eq}), the continuous potential 
$\calU_{\rm pl}$ of a single plane can be presented in the form of a series:
\begin{eqnarray}
\calU_{\rm pl}(a; y,z)
 = 
\calU_0(a; y) 
+
\sum_{m=1}^{\infty} \cos(2\pi mz/\lamu)\,\calU_m(a;y)~.
\label{Pot-CU:eq.10}
\end{eqnarray}
The expansion potentials $\calU_m(a;y)$ can be related to the atomic potentials 
according to the expressions given in Ref.~\cite{Dedkov1994}. 
In the limit $a \to 0$ the expansion potentials with $n>0$ vanish, $\calU_m(0;y)=0$, and 
the zero-order 
(non-periodic in $z$) term yields the continuous potential of the straight plane, 
$\calU_{\rm pl}(0; y,z) = \calU_0(0; y) \equiv \calU_{\rm pl}(y)$. 
The later 
planar potential depends only on the transverse coordinate $y$. 
In the following, we assume the bending amplitude and period to satisfy the 
SASP bending condition 
\begin{equation}
a < d \ll \lamu,
\label{Pot-CU:eq.02}
\end{equation}
and relate the non-periodic term $\calU_0(a; y)$ to the planar potential 
$\calU_{\rm pl}(y)$. 

For a straight plane, the continuous potential is obtained by summing up the atomic 
potentials as exerted by the atoms distributed uniformly along the 
plane~\cite{Lindhard1965}. The surface density $\calN$ of the atoms is related to the 
volume density $n$ as $\calN=nd$, where $d$ is the inter-planar spacing.

\begin{figure} [ht]
\centering
\includegraphics[scale=0.4,clip]{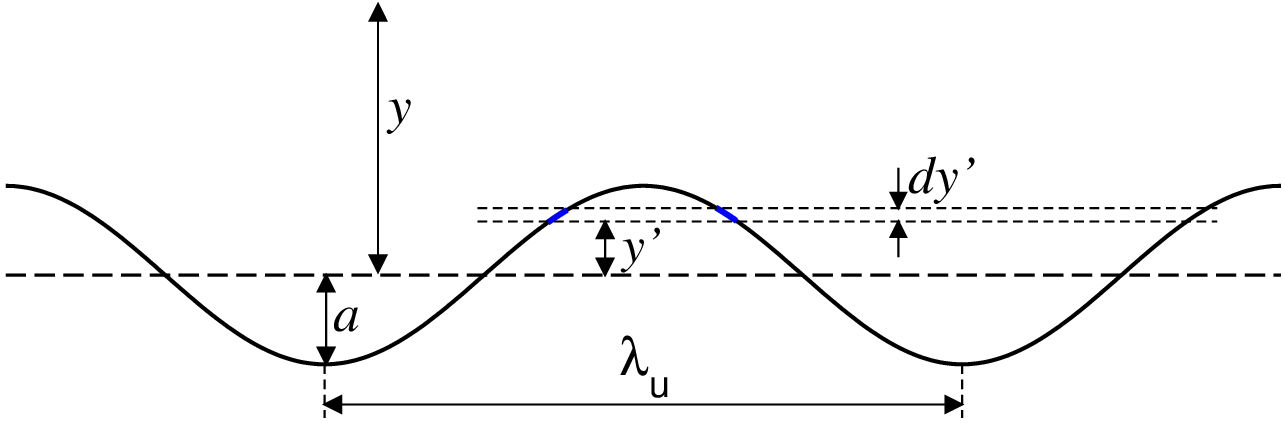}
\caption{
Illustrative figure to the method of calculating the continuous potential for a 
periodically bent crystallographic plane (thick solid curve represents the 
bending profile (\ref{PB_profile:eq})). See explanations in the text.}
\label{fig.A1} 
\end{figure}

For the periodically bent plane we notice, that the atoms located withing the 
interval $[y^{\prime}, y^{\prime}+\d y^{\prime}]$ with respect to the centerline 
$y^{\prime}=0$ are distributed along the planar surface with the density 
$2\d l \calN/\lamu$.   By virtue of the strong inequality $a \ll \lamu$, 
Eq.~(\ref{Pot-CU:eq.02}) above, we have $\d l \approx \d z$ and can approximate 
the planar atomic density as $2\d z \calN/\lamu$ (see Fig. \ref{fig.A1}). 
The continuous potential then can be calculated as follows 
\begin{eqnarray}
\calU_0(a; y) &=& \int_{-\lamu/4}^{\lamu/4}
{2\d z \over \lamu}\, \calU_{\rm pl}\Bigl(\left|y-a\cos(2\pi z/\lamu)\right|\Bigr)
\nonumber \\
              &=& {1\over \pi } \int_{-\pi/2}^{\pi/2}
\calU_{\rm pl}\Bigl(\left|y-a\cos\xi\right|\Bigr) \d \xi~,
\label{Pot-CU:eq.05}
\end{eqnarray}
where $\xi=2\pi z/\lamu$, and appears to not depend on the bending 
period $\lamu$ but be affected by the amplitude $a$. For $a=0$ the potential 
reduces to potential $\calU_{\rm pl}(y)$ for the straight plane. 

Similar procedure can be used to derive a non-periodic part of the distribution 
$n^{(+)}_0(a;y)$ of the nuclei in the periodically bent plane. One obtains  
\begin{equation}
n^{(+)}_0(a;y) = {1\over \pi } \int_{-\pi/2}^{\pi/2}
n_{\rm pl}^{(+)}\Bigl(\left|y-a\cos\xi\right|\Bigr)
\d \xi \,,
\label{density:eq.01}
\end{equation}
where $n_{\rm pl}^{(+)}$ is the nuclear volume density for the straight plane. 
With account for the thermal vibrations, the latter density is given by the 
formula 
\begin{equation}
n_{\rm pl}^{(+)}(y) = {n \over \sqrt{2\pi u_T^2}}\,
\exp\left(-{y^2 \over 2 u_T^2}\right)\,,
\label{density:eq.02}
\end{equation}
where $u_T$ is the root-mean-square amplitude of the vibrations. 
The corresponding non-periodic part of distribution $n^{(-)}_0(a;y)$ of the 
crystalline electrons can be calculated from the Poisson equation as 
\begin{equation}
n^{(-)}_0(a;y) = {1 \over 4\pi e^2 }{\d^2 \calU_0(a;y) \over \d y^2}
+ Z n^{(+)}_0(a;y)\,,
\label{density:eq.03}
\end{equation}
where $Z$ is the nucleus charge.

\subsection{Continuous Inter-planar Potential \label{ContPot:CU}}

The non-periodic part $U_0(a; y)$ of the inter-planar potential is obtained 
by summing up the potentials (\ref{Pot-CU:eq.05}) of the separate planes. 
For the electrons it can be presented in the form 
\begin{equation}
U_0(a; y) = \calU_0(a; y) 
          + \sum_{n=1}^{N_{\max}} 
            \left[ \calU_0(a; y+nd) + \calU_0(a; y-nd) \right]
          + C\,,
\label{ContPot:eq.01}
\end{equation}
where $y$ is the transverse coordinate with respect to an arbitrary selected 
reference plane, and the sum describes a balanced contribution from the 
neighboring planes. Since the planar potential (\ref{Pot-CU:eq.05}) falls off 
rapidly with increasing distance from the plane, Eq.~(\ref{ContPot:eq.01}) 
provides a good approximation for the inter-planar potential at already 
moderate numbers of the terms included in the sum. In our calculations we 
use $N_{\max}=2$. The constant term $C$ can be varied to adjust $U_0(a; y=0)=0$.
For the positrons, the inter-planar potential can be obtained from 
Eq.~(\ref{ContPot:eq.01}) by reversing the signs of the planar potentials and 
selecting the constant $C$ to adjust $U_0(a;y=\pm d/2)=0$. Similar summation 
schemes allow to calculate the charge densities, nuclear and electronic, across 
the periodically bent channels. 

\subsection{The Moli\`{e}re Approximation \label{ContPot:Moliere}}

The integral in the right-hand side of Eq.~(\ref{Pot-CU:eq.05}) 
for the continuous potential for a separate periodically bent plane 
can be evaluated using various model atomic potentials to quantify 
the potential $\calU_{\rm pl}(y)$ for a straight crystalline plane. 
A variety of model potentials can be found in, e.g., 
Refs.~\cite{Lindhard1965,Gemmel,Baier,BiryukovChesnokovKotovBook,Uggerhoj_RPM2005}.
In our studies we use the Moli\`ere approximation for the atomic 
potentials \cite{Moliere} and evaluate the planar potential with 
accounting for the thermal vibrations of the atoms 
(cf. Eq.~(\ref{density:eq.02})). 
This approach yields the planar 
potential in a closed analytical form
\cite{Erginsoy_PRL_v15_360_1965,AppletonEtAl_PR_v161_330_1967}
\begin{eqnarray}
\calU_{\rm pl}(y) = 2\pi n_{\rm am} d\, Z e^2 \aTF
\sum\limits_{i=1}^3 \Bigl( F_i(y) + F_i(-y)\Bigr)
\label{AppendixC.2}
\end{eqnarray}
with
\begin{eqnarray}
F_i(\pm y) =
{\alpha_i \over 2\beta_i} 
\exp\left(
{\beta_i^2u_T^2 \over 2a_{\rm TF}^2} 
\pm 
{\beta_i y \over \aTF}
\right)\,
\mathrm{erfc}\left[{1\over \sqrt{2}} 
\left({\beta_i u_T \over \aTF} \pm {y \over u_T} \right)
\right].
\label{AppendixC.3}
\end{eqnarray}
In the above equations, $n_{\rm am}$ is the mean nuclear density in the amorphous 
medium, $Z$ is charge number of the crystalline atoms, 
$\aTF = 0.8853 a_{\mathrm{B}} Z^{-1/3}$ is the Thomas-Fermi radius 
($a_{\mathrm{B}} = 0.529$ \AA{} is the Bohr radius), 
$\alpha_{1,2,3}=(0.1, 0.55, 0.35)$ and $\beta_{1,2,3}=(6.0, 1.2, 0.3)$ are 
the parameters of the Moli\`ere approximation for the atomic potential. 
The complementary error functions, 
$\mathrm{erfc}(\zeta) = 2\pi^{-1/2}\int_{\zeta}^{\infty} \exp(-t^2)\, \d t$, 
in Eq.~(\ref{AppendixC.3}) result from averaging over the thermal 
vibrations of atomic nuclei.

\begin{figure} [ht]
\centering
\includegraphics[scale=0.31,clip]{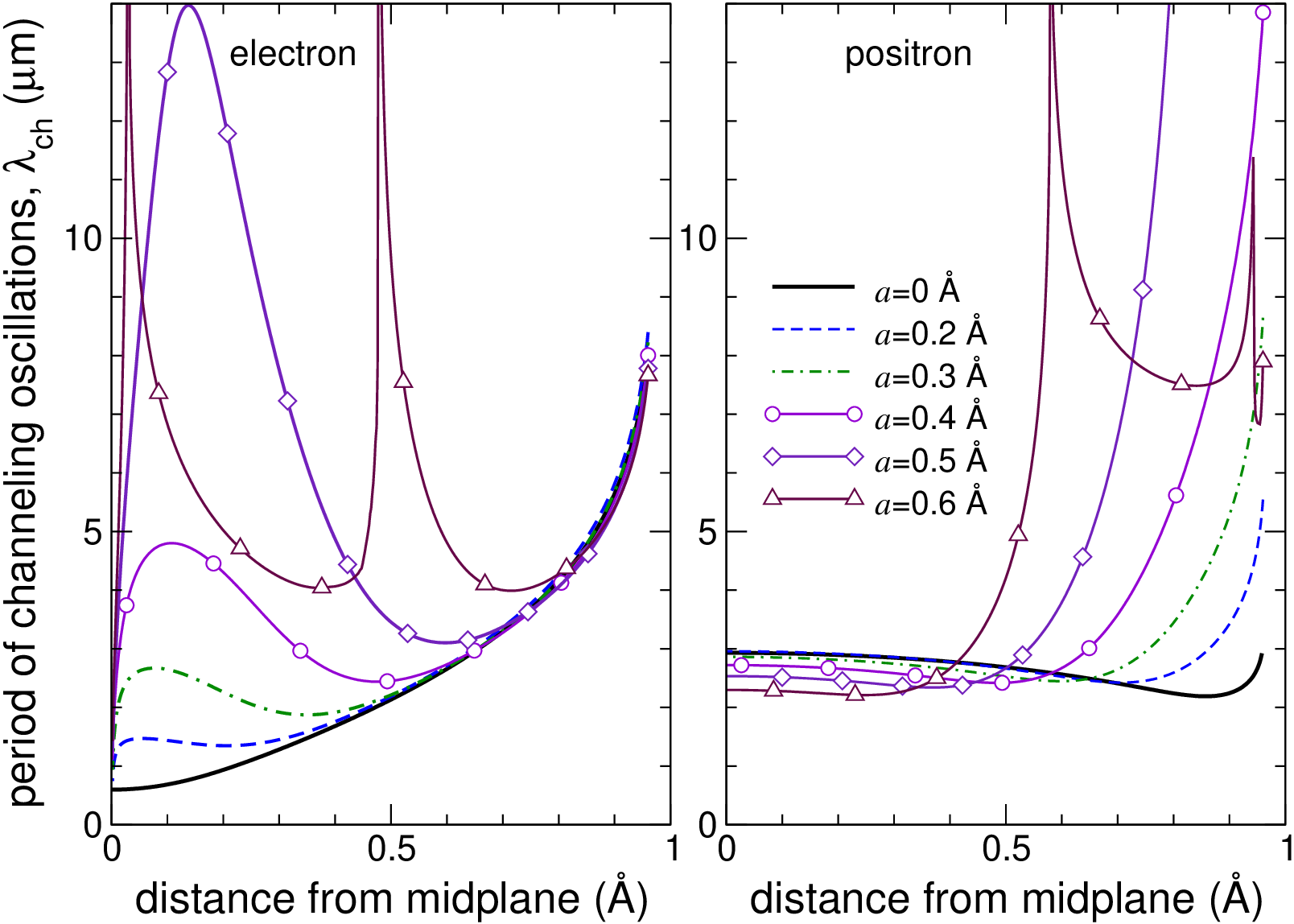}
\\
\includegraphics[scale=0.31,clip]{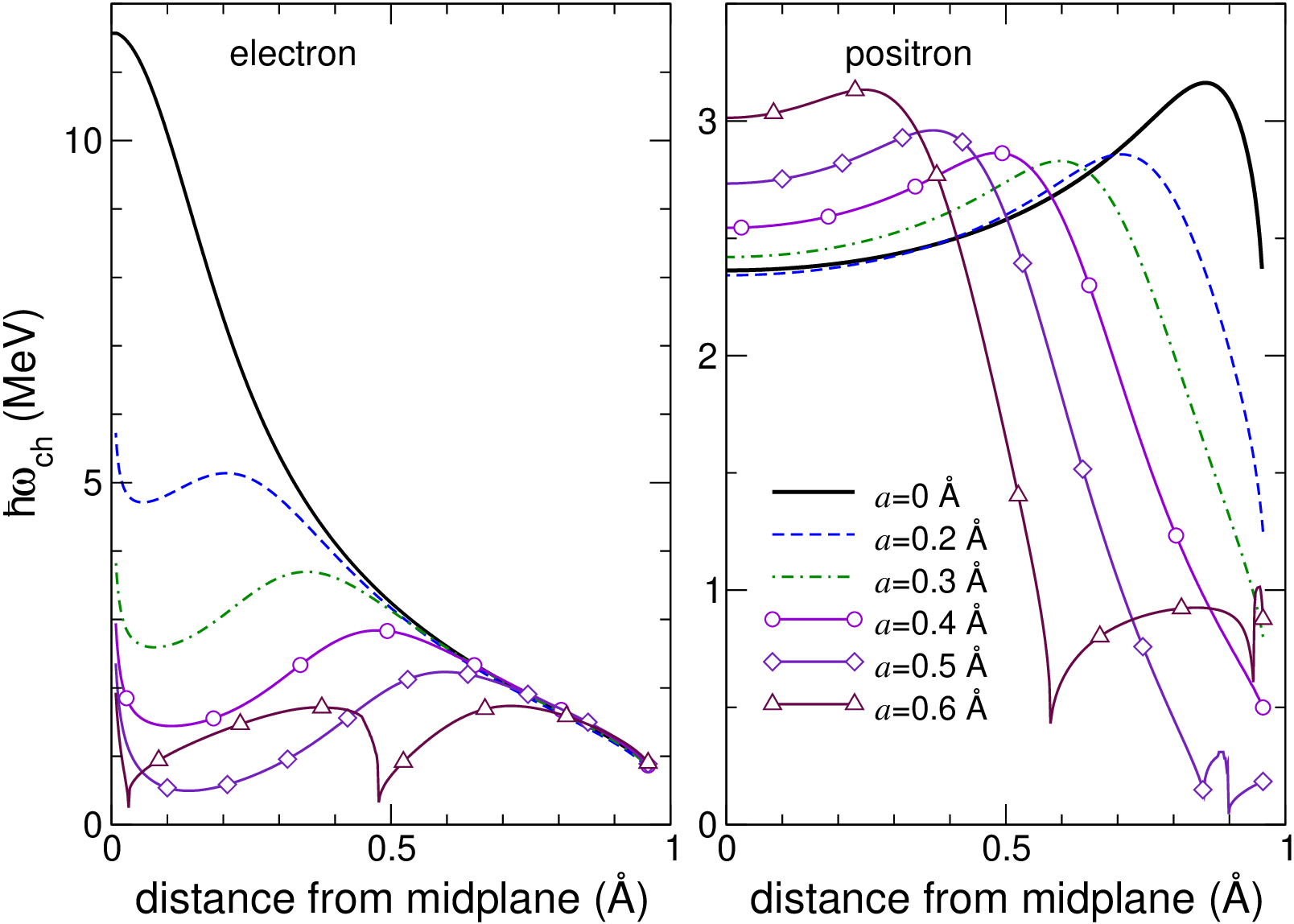}
\caption{
Periods (upper plots) and energies (lower plots) of channeling oscillations for 
855 MeV electrons (left panels) and positrons (right panels). 
The periods and energies are studied as functions of the distance from the channel 
centerlines for the continuous potentials shown in Fig. \ref{fig.03}. 
The curves correspond to different bending amplitudes $a$ as indicated 
($a=0$ corresponds to the straight crystal). See also explanations in the text.
}
\label{fig.A2} 
\end{figure}

The inter-planar potentials, presented in Fig.~\ref{fig.03}, were 
calculated from Eqs. (\ref{Pot-CU:eq.05}), (\ref{ContPot:eq.01}), 
(\ref{AppendixC.2}) and (\ref{AppendixC.3}). These potentials are also helpful in 
understanding the evolution of the lines of channeling radiation with varying 
bending period. To clarify this evolution we have investigated the periods 
$\lambda_{\rm ch}$ of the channeling oscillations in the trajectories 
and the corresponding radiation energies $\hbar\omega_{\rm ch}$. These quantities 
have been evaluated as the functions of the amplitude of channeling trajectory 
oscillations. The amplitudes were fixed by the maximal transverse displacements 
of the projectiles from the channel centerlines, and the periods have been 
evaluated from the corresponding classical turning points and the shapes for the 
continuous inter-planar potential (see, e.g., Sec.~C.2 in 
Ref.~\cite{ChannelingBook2014}). The channeling energies we deduced from the periods 
according to the relation $\omega_{\rm ch} = 2\gamma^2 (2\pi{c}/\lambda_{\rm ch})$. 
The results for both types of the projectiles in the straight and bent with 
different amplitudes channels are shown in 
Fig.~\ref{fig.A2}. For the straight crystals, the channeling 
radiation energies vary with the amplitude of the trajectory oscillations in a narrow 
energy range for the positrons and in a prominently broader range for the electrons 
(see the lower plots in the figure). The latter properties correspond to the narrow 
channeling lines in the radiation spectra for the positrons, and the the broader lines 
in the spectra for the electrons, as seen in Fig.~\ref{fig.05}. 
With changing bending amplitude, the variation ranges of the radiation energy 
increase for the positrons and decrease for the electrons, and correspond to the 
evolution of the channeling lines in the simulated spectra.

\section*{References}

\end{document}